\documentclass{elsarticle}

\usepackage{bm,amssymb,amsmath}
\usepackage{hyperref}

\newcounter{bla}

\newcommand{\manual}[1]{\textit{MANUAL NOTE: #1}}
\renewcommand{\manual}[1]{}
\newcommand{\hippy}{\textsc{HiPPy}}
\newcommand{\hpsrc}{\textsc{HPsrc}}
\newcommand{\vegas}{\textsc{Vegas}}

\newcommand{\bld}[1]{\bm{#1}}
\newcommand{\tr}{\mathop{\mathrm{Tr}}}
\newcommand{\re}{\mathop{\mathrm{Re}}}
\newcommand{\tA}{\tilde{A}}
\newcommand{\half}{\frac{1}{2}}
\newcommand{\ourtitle}{Automated generation of lattice QCD Feynman rules}
\journal{Computer Physics Communications}

\begin{document}
\begin{frontmatter}

\title{\ourtitle}

\author[edinburgh]{A. Hart}
\author[zeuthen]{G.M. von Hippel}
\author[cambridge]{R.R. Horgan}
\author[edinburgh]{E.H. M\"{u}ller}

\address[edinburgh]{SUPA, School of Physics and Astronomy, 
University of Edinburgh, King's Buildings, Edinburgh EH9 3JZ, U.K.}  
\address[zeuthen]{Deutsches Elektronen-Synchrotron DESY,
Platanenallee 6, 15738 Zeuthen, Germany}
\address[cambridge]{DAMTP, CMS, University of Cambridge, Wilberforce Road,
Cambridge CB3 0WA, U.K.}

\begin{abstract}

The derivation of the Feynman rules for lattice perturbation theory
from actions and operators is complicated, especially for highly
improved actions such as HISQ. This task is, however, both important
and particularly suitable for automation. We describe a suite of
software to generate and evaluate Feynman rules for a wide range of
lattice field theories with gluons and (relativistic and/or heavy)
quarks. Our programs are capable of dealing with actions as
complicated as (m)NRQCD and HISQ. Automated differentiation methods
are used to calculate also the derivatives of Feynman diagrams.

\end{abstract}


\end{frontmatter}


{\bf PROGRAM SUMMARY}

\begin{small}
\noindent
{\em Manuscript Title:}
\ourtitle \\
{\em Authors:}
A. Hart, G.M. von Hippel, R.R. Horgan, E.H. M\"{u}ller. \\
{\em Program Title:} 
\hippy, \hpsrc \\
{\em Journal Reference:} \\
%
{\em Catalogue identifier:} \\
%
{\em Licensing provisions:}
GPLv2 (see note in Sec.~\ref{sec_licence}.) \\
{\em Programming languages:}
Python, Fortran95 \\
{\em RAM:}
Problem specific, typically less than 1GB for either code. \\
%
%
{\em Classification:}
4.4 Feynman diagrams; 
11.5 Quantum Chromodynamics, Lattice Gauge Theory \\
{\em Nature of problem:}\\ 
  Derivation and use of perturbative Feynman rules for complicated
  lattice QCD actions.
   \\
{\em Solution method:}\\
An automated expansion method implemented in Python (\hippy) and
code to use expansions to generate Feynman rules in Fortran95
(\hpsrc).
   \\
{\em Restrictions:}\\
No general restrictions. Specific restrictions are discussed in the text.
   \\
%
{\em Running time:}\\
Very problem specific, depending on the complexity of the Feynman
rules and the number of integration points. Typically between a few
minutes and several weeks.
\\
%
\end{small}

\newpage


\hspace{1pc}
{\bf LONG WRITE-UP}

\section{Introduction}

Non--abelian gauge theories are the most important ingredient in our
present understanding of elementary particles and their interactions.
In particular, quantum chromodynamics (QCD) is now universally
believed to be the correct theory of the strong interactions. However,
while perturbation theory has been used successfully in describing the
scattering of particles by partons, the perturbative series does not
converge at hadronic energy scales. Moreover, the phenomena of
confinement and the hadronic spectrum are fundamentally beyond the
reach of perturbation theory. Therefore, non-perturbative Monte Carlo
simulations of lattice-regularised QCD are crucial in order to obtain
a full description and understanding of QCD phenomena.

The lattice regularisation with a lattice spacing $a$ does, however,
introduce a sharp momentum cutoff at the momentum scale $\pi/a$.
Connecting lattice measurements to their continuum counterparts
therefore requires renormalisation factors accounting for the excluded
high-frequency modes. In particular, renormalisation is needed for QCD
matrix elements, and for fixing the bare quark masses to be used in
the lattice Lagrangian. Renormalisation is also necessary to determine
the running of the strong coupling constant $\alpha_s$ and to relate
the lattice regularisation scale $\Lambda_{\mathrm{lat}}$ to
$\Lambda_{\mathrm{QCD}}$. Since the lattice regularisation also
introduces discretisation errors, renormalisations are also used to
``improve'' the lattice action in order to reduce these discretisation
errors at a given lattice spacing.

While in some cases the renormalisation constants can be determined
non-perturbatively
\cite{Martinelli:1994ty,Sommer:2006sj},
results at finite lattice spacing can depend upon
the method used (cf. e.g.
\cite{Bhattacharya:2000pn}),
and non-perturbative methods do not cope well with operators that mix
under renormalisation. For these reasons, lattice perturbation theory
plays an important role in determining the renormalisation constants
needed to extract continuum predictions from lattice QCD.

Given the breakdown of perturbation theory in low energy QCD, one
might doubt whether it could work on the lattice. An argument in
favour of its use is given in
\cite{Lepage:1996jw}:
since the renormalisation factors may be thought of as compensating
for the ultraviolet modes excluded by the lattice regulator, and since
for typical lattice spacings $a \lesssim 0.1 \textrm{~fm}$, the
excluded modes have momenta in excess of 5 GeV, the running QCD
coupling $\alpha_s(\pi/a)$ is small enough that perturbation theory
should rapidly converge. The wide range of results reviewed
in
\cite{Capitani:2002mp,Trottier:2003bw}
show that perturbation theory is useful for a large range of lattice
QCD processes. The assumption that non-perturbative effects do not
contribute on these short length scales, can be tested directly in
some cases by comparing higher order perturbative calculations with
Monte Carlo simulations performed at a range of weak couplings
\cite{Lepage:1999qr,DiRenzo:2000ua,Horsley:2001uy,Trottier:2001vj,Hart:2004jn,Allison:2008ri}),
or by using so-called stochastic techniques
\cite{DiRenzo:2007qf}.
In all these cases, the non-perturbative contributions turned out to be
small. Other comparisons, such as
\cite{Bhattacharya:2000pn},
cannot distinguish non-perturbative effects from higher-order
perturbative corrections. 

Lattice perturbation theory therefore
provides a reliable, and the only systematically improvable, method
for determining the full range of renormalisation constants
\cite{Capitani:2002mp}.

As in the continuum, the calculation of lattice Feynman diagrams is a
two--stage process. First, the lattice action and any operator
insertions are Taylor--expanded in the (bare) coupling constant to
give the propagators and vertices that form the Feynman rules (the
``vertex expansion'' stage). Secondly, these rules are then used to
construct and numerically evaluate Feynman diagrams, possibly after
some algebraic simplification (the ``Feynman diagram evaluation''
stage).

The latter task is more complicated than in the continuum due to the
presence of Lorentz symmetry violating terms at finite lattice
spacing, and on a finite lattice volume also by the more complicated
nature of discrete momentum sums as compared to momentum integrals.
Diagrams are therefore usually evaluated using numerical routines
like \vegas\
\cite{Lepage:1977sw,Lepage:1980dq},
or proprietary mathematical packages, possibly after manipulation
using other computer algebra packages like \textsc{Form}
\cite{Vermaseren:2000nd}.

The greater difficulty is posed, however, by the task of vertex
expansion. Deriving the Feynman results on the lattice is far more
complicated than in the continuum for a number of reasons. Firstly,
lattice gauge fields are elements of the gauge Lie group itself rather
than of its Lie algebra. To obtain the perturbative expansion of the
action, we must therefore expand exponentials of non--commuting
objects. As a consequence, the Feynman rules even for the simplest
lattice action are already much more complicated than their continuum
counterparts.

Secondly, modern lattice theories generally contain a large number of
additional (renormalisation group irrelevant) terms chosen to improve
specific aspects of the Monte Carlo simulation, such as the rate of
approach to the continuum or chiral limits of QCD. Since there is no
unique prescription for these terms, and the best choice depends on
which quantities we are most interested in simulating, a large number
of different lattice actions and operators are in use. Subtle though the
differences between the lattice formulations may be, each choice
provides a completely separate regularisation of QCD with its own set
of renormalisation constants and in particular its own lattice Feynman
rules. For a long time, the complications of the perturbative
expansion have led to the calculations of renormalisation factors
lagging far behind new developments in the improvement of lattice
theories. In many cases this has restricted the physical predictions
that could be obtained from the simulations.

An automated method for deriving lattice Feynman rules for as wide a
range of different theories as possible is therefore highly desirable.
The vertex expansion should be fast enough not to impose undue
constraints on the choice of action. To avoid human error, the user
should be able to specify the action in a compact and intuitive
manner. Since the evaluation of the Feynman diagrams can be
computationally intensive, and will often need to be carried out in
parallel on costly supercomputing facilities, parsimony dictates that
the Feynman rules should be calculated in advance on a different
system, and rendered as machine readable files that can be copied to
the supercomputer for the Feynman diagram evaluation stage.

In this paper we describe a pair of software packages for deriving the
Feynman rules for arbitrary lattice actions%
\footnote{We shall use the term ``action'' so as to include measurement
  operators here and in the following.}
and for evaluating the resulting vertices in a numerical Feynman
diagram calculation. Our algorithm is based through our older
algorithm
\cite{Hart:2004bd}
on the seminal work of L\"{u}scher and Weisz
\cite{Luscher:1985wf}.
A different implementation of the latter has been used in
\cite{Nobes:2001tf,Nobes:2002uu,Nobes:2003nc},
and a similar method is employed in
\cite{Alles:1992yh}.

The new feature of the algorithm presented here is that it is capable
of expanding not only gluonic actions like the algorithm of
\cite{Luscher:1985wf},
and fermionic actions like our algorithm from
\cite{Hart:2004bd},
but also far more complicated multiply-smeared fermionic actions
with reunitarisation such as HISQ
\cite{Follana:2006rc},
and that it supports taking advantage of the factorisation inherent
in some lattice actions, such as improved lattice formulations of NRQCD
\cite{Lepage:1992tx}.

As in
\cite{Luscher:1985wf} and \cite{Hart:2004bd},
the vertex expansion is performed completely independently of any
boundary conditions, allowing for instance, the use of twisted
periodic boundary conditions as a gauge--invariant infrared regulator
\cite{Luscher:1984xn,Luscher:1985zq}
or for changing the discrete momentum spectrum in other ways
\cite{deDivitiis:2004kq}.

We have used the software packages described for calculations of the
renormalised anisotropy in gauge theories
\cite{Drummond:2002yg,Drummond:2003qu},
to study the mean link in Landau gauge for tadpole improvement
\cite{Hart:2004jn},
to measure the electromagnetic decays of heavy quark systems using
NRQCD
\cite{Drummond:2002kp,Hart:2006ij,Hart:2006uj,Hart:2007zza},
to calculate the radiative corrections to the gluonic action due to
Highly Improved Staggered (HISQ) sea quarks
\cite{Hart:2008zi,Hart:2008sq}
and the renormalisation of the self energy of heavy quarks using
moving NRQCD (mNRQCD)
\cite{Meinel:2008th}.

The code is flexible and can be easily extended, as has already been
done for lattice--regularised chiral perturbation theory
\cite{Borasoy:2005nz},
perturbative calculations in the Schr\"{o}dinger functional
\cite{Takeda:2007dt,Takeda:2008rr}
and anisotropy calculations
\cite{Foley:2008mp}.

The structure of this paper is as follows. In the next Section, we
review the basic expansion algorithm described in
Ref.~\cite{Hart:2004bd}, outlining some improvements in, for instance,
the handling of automatic derivatives and spin matrices. In
Sec.~\ref{sec_new_stuff}, we present novel extensions to the algorithm
that are needed to describe complicated fermion actions, including
HISQ, NRQCD and mNRQCD.

Sec.~\ref{sec_software} provides details of the implementation of the
algorithm in the \hippy\ and \hpsrc\ codes and of their installation,
testing and use. We make some concluding remarks in
Sec.~\ref{sec_conc}. Technical details are relegated to the
appendices.

\subsection{Licence}
\label{sec_licence}

The \hippy\ and \hpsrc\ codes are released under the
second version of the GNU General Public Licence (GPL v2).
Therefore anyone is free to use or modify the code for their own
calculations.

As part of the licensing, we ask that any
publications including results from the use of this code or of
modifications of it cite Refs.~\cite{Luscher:1985wf,Hart:2004bd}
as well as this paper.

Finally, we also ask that details of these publications, as well as of
any bugs or required or useful improvements of this core code, would be
communicated to us.

\section{Theoretical background}
\label{sec_old_stuff}

In this section we describe the algorithms used to give the most
efficient, yet generic, implementation of the Feynman rules for
lattice actions. These extend the original work of L\"uscher and Weisz
\cite{Luscher:1985wf} and developments described in
Ref.~\cite{Hart:2004bd}.
\subsection{Fields on the lattice}
\label{sec_algorithm}

We consider a $D$-dimensional hypercubic spacetime lattice
with lattice spacing $a$ and extent $L_\mu a$ in the $\mu$-direction:
\begin{equation}
\Lambda = \left\{(x_1,\ldots,x_D)\in\mathbb{R}^D ~~\middle|
\; \forall \; \mu \in \{1,\ldots,D\} :~
                          \frac{x_\mu}{a} \in \{0,\ldots,L_\mu-1\} 
\right\}
\end{equation}
where lattice sites are labelled by a vector $\bld{x} \in \Lambda$. In
the following, we will usually set $a=1$ (a lattice anisotropy can be
introduced by rescaling the coupling constants in the action
\cite{Drummond:2002yg}).
Let $\bld{e}_\mu$ be a right-handed orthonormal basis,
and $\bld{e}_{-\mu} \equiv -\bld{e}_\mu$.

\manual{In the \hippy\ code, the dimension is specified in 
\texttt{parameters.D}. In the \hpsrc\ code, it is given by 
\texttt{Ndir} in \texttt{mod\_phys\_parm.F90}. Note that neither 
code has been fully tested for $D \neq 4$}

A lattice path consisting of $l$ links starting at site $\bld{x}$ can
be specified by an ordered set of directions given by integers, $s_i
\in
\{-D,\ldots,-1,1,\ldots,D\}$:
\begin{equation}
\mathcal{L}(\bld{x},\bld{y};\bld{s}) \equiv \{ 
\bld{x},\bld{y}; \bld{s} = [s_0,s_1,\ldots,s_{l-1}] \} \;,
\label{eqn_path_def}
\end{equation}
with the $j^{\, \mathrm{th}}$ point on the path being
\begin{equation}
\bld{z}_j = \left\{
\begin{array}{ll}
\bld{x} \;, & ~~j = 0 \; ,
\\
\bld{z}_{j-1} + a \bld{e}_{s_{j-1}} \; , &
~~ 0 < j \leq l \; ,
\end{array}
\right.
\label{path}
\end{equation}
and $\bld{y} \equiv \bld{z}_l$. 

For periodic boundary conditions, the momentum vectors are
\begin{equation}
\bld{k} = \frac{2\pi}{a} \, 
\left( \frac{\bar{k}_1}{L_1}, \ldots, \frac{\bar{k}_D}{L_D} \right) \; ,
~~~~
0 \le \bar{k}_\mu < L_\mu \; , 
~~~~
\bar{k}_\mu \in \mathbb{Z} \; ,
\end{equation}
and the Fourier expansion of a function $\phi$ is
\begin{equation}
\tilde{\phi}(\bld{k}) = \sum_{\bld{x}} \, 
e^{ -i \bld{k} \cdot \bld{x} } \phi(\bld{x}) \;,
~~~~~~
\phi(\bld{x})
= \frac{1}{V} \sum_{\bld{k}} \,
e^{ i \bld{k} \cdot \bld{x} } \tilde{\phi}(\bld{k}) \; ,
\end{equation}
where $V=\prod_\mu L_\mu$ is the lattice volume.

Twisted boundary conditions
\cite{'tHooft:1979uj}
provide a useful gauge-invariant infrared regulator in perturbative
calculations
\cite{Luscher:1985wf}.
These change the momentum spectrum, converting colour factors into
``twist matrices'' associated with momenta interstitial to the
reciprocal lattice (with the introduction of an additional quantum
number, ``smell'', for fermions
\cite{Parisi:1984cy,Wohlert:1987rf,Hart:2004jn}).
The \hpsrc\ code fully supports such boundary conditions but for
simplicity we only discuss periodic boundary conditions in this paper.

\manual{In the \hpsrc\ code, the number of directions with 
twisted boundary conditions (0, 2, 3 or 4) is specified by 
\texttt{ntwisted\_dirs} in \texttt{mod\_phys\_parm.F90}. 
Clebsch-Gordan factors for all cases are cases are calculated 
in \texttt{mod\_colours.F90}.}

Following
Ref.~\cite{Luscher:1985wf},
we denote the gauge field associated with a link as $U_{\mu > 0}(\bld{x}) \in
SU(N)$, and define $U_{-\mu}(\bld{x}) = U^\dagger_\mu(\bld{x} - a
\bld{e}_\mu)$.
The gauge potential $A_\mu\in\mathop{\mathrm{alg}}(SU(N))$ associated with
the midpoint of the link is defined through
\begin{equation}
U_{\mu>0}(\bld{x}) = \exp \left(a g A_\mu \left( 
\bld{x} + \frac{a}{2} \bld{e}_\mu \right) \right) = 
\sum_{r=0}^\infty \frac{\left( ag A_\mu(\bld{x} + \frac{a}{2} \bld{e}_\mu) 
\right)^r}{r!}
\label{eqn_link_exp}
\end{equation}
where $g$ is the bare coupling constant.
In terms of the anti-Hermitian generators of $SU(N)$,
\begin{equation}
A_\mu = A_\mu^a\,T_a, \qquad
\left[ T_a,T_b \right] = -f_{abc} T_c, \qquad
\tr \left( T_a T_b \right) = -\half \, \delta_{ab} \;.
\end{equation}
Quark fields $\psi(\bld{x})$ transform according to the
representation chosen for the generators $T_a$, which we take
to be the fundamental representation (other choices will affect
the colour factors, but not the structure of our algorithm).

\subsection{Perturbative expansion of Wilson lines}

The Wilson line $L(\bld{x},\bld{y},U)$ associated with the
lattice path $\mathcal{L}(\bld{x},\bld{y};\bld{s})$ is a product of links
\begin{equation}
L(\bld{x},\bld{y},U) = 
\prod_{i=0}^{l-1} U_{s_i}(\bld{z}_i) 
= \prod_{i=0}^{l-1} \exp \left[ \mathop{\mathrm{sgn}}(s_i) a 
g A_{|s_i|} \left( \bld{z}_i+\frac{a}{2}\bld{e}_{s_i} \right) \right] \; . 
\end{equation}
As all actions and operators can be written as sums of Wilson lines
(possibly terminated by fermion fields), our goal is to efficiently
expand $L$ in terms of the gauge potential in momentum space:
\begin{multline}
L(\bld{x},\bld{y} ; A) = \sum_r \frac{(ag)^r}{r!}
\sum_{\bld{k}_1,\mu_1,a_1} \ldots
\sum_{\bld{k}_r,\mu_r,a_r}
\tilde{A}_{\mu_1}^{a_1}(\bld{k}_1) \ldots
\tilde{A}_{\mu_r}^{a_r}(\bld{k}_r) \times
\\
V_r(\bld{k}_1,\mu_1,a_1 ; \ldots ; \bld{k}_r,\mu_r,a_r) \; .
\label{eqn_tayl_exp}
\end{multline}
The vertex functions $V_r$ factorise as
\begin{equation}
V_r(\bld{k}_1,\mu_1,a_1 ; \ldots ; \bld{k}_r,\mu_r,a_r) = 
C_r(a_1, \ldots , a_r) \;
Y^{\mathcal{L}}_r(\bld{k}_1,\mu_1 ; \ldots ; \bld{k}_r,\mu_r)
\end{equation}
with a momentum- and path-independent Clebsch--Gordan
(colour) factor $C_r$
\begin{equation}
C_r(a_1, \ldots , a_r) = \prod_{i=1}^{r} T_{a_i} \; .
\end{equation}
It is therefore more efficient to calculate just the expansion of the
reduced vertex functions, $Y^{\mathcal{L}}_r$ (with an appropriate
description of the colour trace structure where ambiguous --- see
Appendix~B of
Ref.~\cite{Hart:2004bd} 
for further details). The reduced vertex function can be written as a
sum of terms, each of which contains an exponential. For convenience,
we will call each term a ``monomial'':
\begin{equation}
Y^{\mathcal{L}}_r(\bld{k}_1,\mu_1 ; \ldots ; \bld{k}_r,\mu_r) = 
\sum_{n=1}^{n_r(\{\mu\})} f^{(r,\{\mu\})}_n \,
\exp \left( \frac{i}{2}  \sum_{j=1}^r
\bld{k}_j \cdot \bld{v}^{(r,\{\mu\})}_{n,j} \right) \; ,
\label{eqn_y}
\end{equation}
where for each combination of $r$ Lorentz indices we have $n_r$ terms,
each with an amplitude $f$ and locations $\bld{v}$ of the $r$ factors
of the gauge potential, which are drawn from the locations of the
midpoints of the links in the path $\mathcal{L}$. To avoid floating
point ambiguities, we express the components of all position
$D$-vectors as integer multiples of $\frac{a}{2}$ (accounting for the
factor of $\half$ in the exponent).

In the \hpsrc\ code, we use the convention that all momenta flow into
the vertex, so $\sum_{i=1}^r \bld{k}_i = 0$.

Eqn.~(\ref{eqn_y}) makes clear that the number of monomials depends
not just on the number of gluons $r$, but also on the choice of
Lorentz indices $\{ \mu \}$, and that each monomial has a different
amplitude and set of $r$ positions. For clarity of presentation, we
will, however, suppress these additional arguments in later
expressions (notably
Eqns.~(\ref{eqn_sy},~\ref{eqn_gy},~\ref{eqn_diff_y},~\ref{eqn_zy},~\ref{eqn_xy})).

\subsubsection{Implementation notes}

The generation of the Feynman rules for generic momenta thus reduces
to a calculation of the amplitudes $f$ and locations $\bld{v}$ of
the monomials that build up the various reduced vertices $Y_r$.

This is all carried out in the \hippy\ code, a description of which
can be found in Sec.~4 of Ref.~\cite{Hart:2004bd}. The amplitudes and
locations defining each monomial are encoded as an instance of the
class \texttt{Entity}, and the collection of monomials that make up
the reduced vertices is encoded as an instance of class
\texttt{Field}. The data structures have been chosen to ensure that
equivalent monomials are combined to minimise the size of the reduced
vertex description.

Once expanded, the monomials required for the reduced vertices at each
order are written to disk as a text file.

\manual{In the \hippy\ code, class \texttt{Entity} is defined in file 
\texttt{class\_entity.py} and class \texttt{Field} is defined in file 
\texttt{class\_field.py}.}

The \hpsrc\ code reads these (previously generated) vertex files at
runtime. For given momenta $\{\bld{k}\}$, lorentz indices $\{\mu\}$
and colour indices, the $Y_r$ are constructed as given in
Eqn.~(\ref{eqn_y}), which is then multiplied by the appropriate
Clebsch-Gordan colour factor(s) to form the (Euclidean) Feynman rule,
$-V_r$.

\subsection{Realistic actions: the fermion sector}

Realistic lattice fermion and gauge actions require some refinements
to this generic description. We begin with the fermion sector. The
most general gauge- and translation-invariant action can be written as
\begin{equation}
S_F(\psi,U) = \sum_{\bld{x}} 
\sum_{\mathcal{W}} h_{\mathcal{W}} \, \bar{\psi}(\bld{x}) \,
\Gamma_{\mathcal{W}} \, W(\bld{x},\bld{y},U) \, \psi(\bld{y})
\end{equation}
and consists of Wilson lines $W$ defined by open paths
$\mathcal{W}(\bld{x},\bld{y};\bld{s})$, each carrying an associated
coupling constant $h_{\mathcal{W}}$ and a spin matrix
$\Gamma_{\mathcal{W}}$ (possibly the identity).

Using the convention that all momenta flow into the vertex, the
perturbative expansion is
\begin{multline}
S_F(\psi,A) = \sum_r \frac{g^r}{r!} 
\sum_{\bld{k}_1,\mu_1,a_1} \ldots
\sum_{\bld{k}_r,\mu_r,a_r}
\tA^{a_1}_{\mu_1}(\bld{k}_1) \ldots 
\tA^{a_r}_{\mu_r}(\bld{k}_r) \times
\\
\sum_{\bld{p},\bld{q},b,c} \tilde{\bar{\psi}}^b(\bld{p}) \,
V_{F,r}(\bld{p},b ; \,\bld{q},c ; 
\, \bld{k}_1,\mu_1,a_1; 
\, \ldots ; 
\, \bld{k}_r,\mu_r,a_r) \, \tilde{\psi}^c(\bld{q}) \;.
\label{eqn_SS}
\end{multline}
The Euclidean Feynman rule for the $r$-point
gluon--fermion--anti-fermion vertex is $-g^r V_{F,r}$, where the
symmetrised vertex is:
\begin{multline}
V_{F,r}(\bld{p},b ; \bld{q},c ; \bld{k}_1,\mu_1,a_1; \, \ldots ; 
\bld{k}_r,\mu_r,a_r)
= 
\\
\frac{1}{r!} \sum_{\sigma \in \mathcal{S}_r}
\sigma \cdot C_{F,r}(b,c;a_1,\ldots,a_r) ~
\sigma \cdot Y_{F,r}(\bld{p},\bld{q} ; 
\bld{k}_1,\mu_1; \, \ldots ; \bld{k}_r,\mu_r) \; ,
\label{eqn_ferm_symm}
\end{multline}
where $\mathcal{S}_r$ is the permutation group of $r$ objects and
$\sigma\in\mathcal{S}_r$ is applied to the gluonic variables,
$\{\bld{k}\}$, $\{\mu\}$ and $\{a\}$. The normalisation factor of $r!$
for this is additional to the $r!$ factor arising from the Taylor
expansion of the exponential in Eqn.~(\ref{eqn_SS}). The reduced
vertex $Y_{F,r} =
\sum_{\mathcal{W}} h_{\mathcal{W}} Y^{\mathcal{W}}_{F,r}$ is the sum
of contributions from paths $\mathcal{W}$.

In most cases the Clebsch-Gordan colour factor is the matrix
element:
\begin{equation}
C_{F,r}(b,c;a_1,\ldots,a_r) = (T_{a_1} \ldots T_{a_r})_{bc} \;,
\end{equation}
and the reduced vertex function has the structure:
\begin{multline}
Y_{F,r}(\bld{p},\bld{q} ; 
\bld{k}_1,\mu_1;\ldots;\bld{k}_r,\mu_r) = 
\sum_{n=1}^{n_r(\{\mu\})} \Gamma_n \,
f_n  \times
\\
\exp \left( \: \frac{i}{2} \left(
\bld{p} \cdot \bld{x} + \bld{q} \cdot \bld{y} +
\sum_{j=1}^r \bld{k}_j \cdot \bld{v}_{n,j} \right) \right) \; ,
\label{eqn_sy}
\end{multline}
where we understand $\Gamma_n \equiv \Gamma_{r,\{\mu\},n}$.
Cases with more complicated colour structures do arise, for example
the use of traceless field strengths in QCD. Such structures are
accommodated in the codes; monomials with different colour structures
are distinguished using ``pattern lists'' (discussed in Appendix~B of
Ref.~\cite{Hart:2004bd}) 
and appropriate colour factors are applied to each when the Feynman
rule is constructed.

As there are no permutation symmetries in $C_{F,r}$, there is no
advantage to carrying out any symmetrisation in the \hippy\ expansion
code. In the \hpsrc\ code, symmetrisation of the Feynman rule shown in
Eqn.~(\ref{eqn_ferm_symm}) carries a potentially significant
computational overhead: the reduced vertices must be calculated afresh
for each permutation. Not all such permutations may be needed because
symmetries of a Feynman diagram can reduce the number of distinct
contributions to its value from the terms in
Eqn.~(\ref{eqn_ferm_symm}). For this reason, symmetrisation is not
carried automatically in the \hpsrc\ code and the user must therefore
explicitly construct all permutations requiring a different
calculation from Eqn.~(\ref{eqn_ferm_symm}), applying the appropriate
normalisation factor.

\subsection{Realistic actions: the gluon sector}

A typical gluonic action is 
\begin{equation}
S(\psi,U) = \sum_{\bld{x}} 
\sum_{\mathcal{P}} c_{\mathcal{P}} \re \tr \left[ 
P(\bld{x},\bld{x},U) \right]  \; ,
\label{eqn_gen_glue}
\end{equation}
built of Wilson loops $P$ defined by closed paths
$\mathcal{P}(\bld{x},\bld{x};\bld{s})$, each with coupling constant
$c_{\mathcal{P}}$. The perturbative action is
\begin{multline}
S_G(A) = \sum_r \frac{g^r}{r!} 
\sum_{\bld{k}_1,\mu_1,a_1} \ldots \sum_{\bld{k}_r,\mu_r,a_r}
\tA^{a_1}_{\mu_1}(\bld{k}_1) \ldots 
\tA^{a_r}_{\mu_r}(\bld{k}_r) \times
\\
V_{G,r}(\bld{k}_1,\mu_1,a_1; \, \ldots ; \, \bld{k}_r,\mu_r,a_r) \; .
\end{multline}
The Euclidean Feynman rule for the $r$-point gluon vertex function is
$(- g^r V_{G,r})$, and the vertex $V_{G,r}$ is
\cite{Luscher:1985wf}
\begin{multline}
V_{G,r}(\bld{k}_1,\mu_1,a_1; \, \ldots ; \bld{k}_r,\mu_r,a_r)
= 
\\
\frac{1}{r!} \sum_{\sigma \in \mathcal{S}_r}
\sigma \cdot C_{G,r}(a_1,\ldots,a_r) ~
\sigma \cdot Y_{G,r}(\bld{k}_1,\mu_1; \, \ldots ; \bld{k}_r,\mu_r) \; ,
\label{eqn_symmetrise}
\end{multline}
The reduced vertex $Y_{G,r} = \sum_{\mathcal{P}} c_{\mathcal{P}}
Y_{G,r}^{\mathcal{P}}$ is the sum of contributions from paths
$\mathcal{P}$. As before, the $(r!)$ factor normalises the
symmetrisation.  $Y^{\mathcal{P}}_{G,r}$ can be expanded as
\begin{equation}
Y^{\mathcal{P}}_{G,r}(\bld{k}_1,\mu_1;\ldots;\bld{k}_r,\mu_r) = 
\sum_{n=1}^{n_r}
f_n  \; \exp \left( \: \frac{i}{2} \sum_i 
\bld{k}_i \cdot \bld{v}_{n,i} \right) \; .
\label{eqn_gy}
\end{equation}
In most cases we expect the lattice action to be real. Thus, for every
monomial $(f_n; \{ \bld{v}_{n,i} \})$ in Eqn.~(\ref{eqn_gy}),
there must be a corresponding term
$((-1)^r f_n^*; \{ -\bld{v}_{n,i} \})$.
We can therefore speed up the evaluation of the Feynman rules by
removing the latter term, and replacing the exponentiation in
Eq.~(\ref{eqn_gy}) with ``$\cos$'' for $r$ even, and with ``$i \sin$''
for $r$ odd. This can either be done by recognising conjugate contours in
the action (e.g. $S = \half \tr [P + P^\dagger ]$) and expanding only
one, or by attaching a flag to each monomial to signal whether its complex
conjugate has already been removed.

If in addition the action has the form Eq.~(\ref{eqn_gen_glue})
with a single trace in the fundamental representation, the colour factors are
\begin{equation}
C_{G,r}(a_1,\ldots,a_r) = \half \left[ \tr \; ( T_{a_1} \ldots T_{a_r}) + 
(-1)^r \tr \; ( T_{a_r} \ldots T_{a_1}) \right] \; .
\label{cg_notwist}
\end{equation}
which has a number of permutation symmetries:
\begin{equation}
\sigma \cdot C_{G,r} = \chi_r(\sigma) \; 
C_{G,r} \; ,\mbox{ where } 
\chi_r(\sigma) = 
\begin{cases}
1 &      \mbox{for $\sigma$ a cyclic permutation,} \\
(-1)^r & \mbox{for $\sigma$ the inversion.}
\end{cases}
\label{eqn_clebsch_sym}
\end{equation}
There is thus a great advantage in carrying out some of the
symmetrisation in Eqn.~(\ref{eqn_symmetrise}) at the expansion stage
in the \hippy\ code. Many of the extra monomials generated by
symmetrising over subgroup $\mathcal{Z}_r$ (generated by cyclic
permutations and inversion) are equivalent and can be combined in the
\hippy\ code, significantly reducing the number of exponentiation
operations required to construct the partially-symmetrised
$Y^\prime_{G,r}$:
\begin{multline}
V_{G,r}(\bld{k}_1,\mu_1,a_1; \, ... ; \,\bld{k}_r,\mu_r,a_r)
= \sum_{\sigma \in \mathcal{S}_r/\mathcal{Z}_r}
\sigma \cdot C_{G,r}(a_1,... ,a_r) \times
\\
\sigma \cdot Y_{G,r}^\prime (\bld{k}_1,\mu_1;\, ... ;\,\bld{k}_r,\mu_r) \; ,
\end{multline}
\begin{equation}
Y_{G,r}^\prime = 
\sum_{\stackrel{\mathcal{P}}{\sigma \in \mathcal{Z}_r}} \; 
c_{\mathcal{P}}
\chi_r(\sigma) \: \sigma \cdot Y^{\mathcal{P}}_{G,r} \; .
\label{v_notwist}
\end{equation}
The $\chi_r(\sigma)$ factors go into the amplitudes
of the new monomials coming from the partial symmetrisation.

The number of symmetrisation steps remaining to be carried out in the
\hpsrc\ code is the number of cosets in
$\mathcal{S}_r/\mathcal{Z}_r$ (one for $r \le 3$, three for $r=4$,
twelve for $r=5$ etc.). These symmetrisation steps (and the
normalisation) are carried out automatically in the \hpsrc\ gluon
vertex modules.

\subsection{Diagram differentiation}
\label{sec_taylor}

In many cases, such as when computing wavefunction renormalisation
constants, one needs to calculate the derivative of a Feynman diagram
with respect to one or more momenta. Whilst derivatives can be
computed numerically using an appropriate local difference operator,
such differencing schemes are frequently numerically unstable and
require computing the Feynman diagram multiple times.
Automatic differentiation methods
\cite{ref_autodiff}
are a stable and cost-saving alternative.

We can easily construct the differentiated Feynman vertex using
Eqn.~(\ref{eqn_y}). If we want to differentiate with repsect to
momentum component $q_\nu$, we first construct a rank $r$ object
$\bm{\tau} = [\tau_1, \ldots, \tau_r]$ which represents the proportion
of momentum $\bm{q}$ in each leg of the Feynman diagram. Momentum
conservation dictates $\sum_i \tau_i = 0$. For instance, for a gluon
3-point function with incoming momenta
$(\bm{p},-\bm{p}+2\bm{q},-2\bm{q})$, we would have $\bm{\tau} =
[0,2,-2]$. The differentiated vertex is
\begin{equation}
\frac{d}{dq_\nu} 
Y^{\mathcal{L}}_r(\bm{k}_1,\mu_1 ; \ldots ; \bm{k}_r,\mu_r) = 
\sum_{n=1}^{n_r} 
\frac{if_n}{2} 
\left( \sum_{j=1}^r \tau_j v_{n,j;\nu} \right)
\exp \left( \frac{i}{2} \sum_{j=1}^r 
\bm{k}_j \cdot \bm{v}_{n,j} \right)
\label{eqn_diff_y}
\end{equation}
and so on for higher derivatives. We may therefore simultaneously
calculate as many differentials as we need for the cost of just one
exponentiation. If this momentum expansion is placed into an
appropriate data structure for which appropriately overloaded operations
have been defined, it is straightforward to create the Taylor series for
a Feynman diagram by simply multiplying the vertex factors together.
We use a slightly modified version of the TaylUR package
\cite{vonHippel:2005dh,vonHippel:2007xd}
to do this, which encodes the Taylor series expansion in the \hpsrc\
Fortran code as a derived type \texttt{taylor}, for which all
arithmetic operations and elementary functions have been overloaded
so as to respect Leibniz's and Fa\`a di Bruno's rules for higher
derivatives of products and functions.

In calculations that require only certain higher-order derivatives,
the taylor multiplication can be significantly sped up by defining a
mask that only propagates certain terms in the Taylor expansion. This
has not, however, been implemented in the distributed version of the
code.

\subsection{Spin algebra}
\label{sec_spinor}

The Feynman rules for fermions contain spin matrices (which may be
Pauli matrices, e.g. for NRQCD, or Dirac matrices for relativistic
fermions). We can expand a generic spin matrix $W$ using a basis
$\{\Gamma_i\}$:
\begin{equation}
W=\sum_{i\in I(W)} w_i \Gamma_i
\end{equation}
The product of any two spin basis matrices $\Gamma_i$ and $\Gamma_j$
is another basis matrix (with label $n_{ij}$) times a phase:
\begin{equation}
\Gamma_i \Gamma_j = \phi_{ij} \Gamma_{n_{ij}} \;,
\label{eqn_spin_mult}
\end{equation}
We choose the basis so that these phases $\phi_{ij}$ are real. Where
another convention is desired, the amplitudes $w_i$ need to be
adjusted appropriately.

For Pauli matrices, we use a basis $\{\Gamma_i\}=\{1,i \sigma_k\}$
and $I(W)\subseteq \{0,\ldots,3\}$, giving:
\begin{equation}
i \sigma_j . i \sigma_k = \begin{cases}
1 & j = k, \\
\epsilon_{jkm} \; i \sigma_m & j \neq k.
\end{cases}
\end{equation}
For the (Euclidean) Dirac matrices, we choose
$\{\Gamma_i\}=\{1,\gamma_\mu,\sigma_{\mu\nu},\gamma_5\gamma_\mu,\gamma_5\}$
and $I(W)\subseteq \{0,\ldots,15\}$. We define $\gamma_5 \equiv
\gamma_1 \gamma_2 \gamma_3 \gamma_4$ and note the definition here
$\sigma_{\mu\nu} \equiv \frac{1}{2} [ \gamma_\mu,\gamma_\nu]$. The
multiplication for the basic $\gamma$ matrices is thus
\begin{equation}
\gamma_\mu \gamma_\nu = \begin{cases}
1 & \mu = \nu, \\
\sigma_{\mu\nu} & \mu \neq \nu.
\end{cases}
\end{equation}
We can thus write the product of two general spin matrices as
\begin{align}
W W' & = \left(\sum_{j\in I(W)} w_j \Gamma_j\right)
       \left(\sum_{k\in I(W')} w'_k \Gamma_k\right)
\nonumber \\
     & = \sum_{i\in I(WW')} \left(\sum_{
\genfrac{}{}{0pt}{2}{j\in I(W),k\in I(W')}{ 
n_{jk}=i}} w_j w^\prime_k \phi_{jk}\right) \Gamma_i
\label{eqn_spin_multiply}
\end{align}
where $I(WW')=\{n_{ij}~|~i\in I(W),~j\in I(W')\}$. 

We use this implicit representation of the spin matrices. Matrix
operations on explicit Dirac matrices take $\mathcal{O}(4^3)$
operations and introduce additional rounding errors.
Eqn.~(\ref{eqn_spin_multiply}), by contrast, needs $|I(W)|\times|I(W')|$
multiplications, depending on how many basis matrices are needed to
describe $W$ and $W^\prime$. If $|I(W)|,|I(W')|<8$, the latter method
is more efficient and this is almost always the case.

There are greater gains when inverting spin matrices of the form 
\begin{equation}
S^{-1} = a_0 1 + \sum_{\mu=1}^4 a_\mu \gamma_\mu \; ,
\label{eq_specialspinor}
\end{equation}
as we might do when obtaining the propagator of a relativistic quark
(for NRQCD, the propagator is spin diagonal in the Pauli matrices and
trivial to invert). The inverse is
\begin{equation}
S = b_0 1 - \sum_{\mu=1}^4 b_\mu \gamma_\mu\;, ~~~~~ 
b_i = a_i\left(a_0^2-\sum_{\mu=1}^4 a_\mu^2\right)^{-1} \; ,
\label{eq_specialspinor_invert}
\end{equation}
which is far more efficient than inverting a $4\times 4$ matrix.
Inversion of a general spin matrix (not of the form
Eqn.~(\ref{eq_specialspinor})) is less efficient with an implicit
representation, but this is irrelevant in most perturbative
calculations.

Since all basis matrices except the identity are traceless, taking the
trace of a spin matrix is a free operation in the implicit
representation.

\subsubsection{Implementation notes}

In the \hippy\ code, spin basis matrices are associated with monomials
using an appropriate integer $i$, which is part of the \texttt{Entity}
data structure. When terms are multiplied together, the factors
$\phi_{jk}$ are absorbed into the amplitude $f$ of the resulting
monomial.

\manual{For efficiency reasons, spin multiplication is not 
automically implemented; multiplication of entities currently
concatenates lists of spin indices. To reduce this list to a single
spin index using the multiplication table, the user must explicitly
call the method
\texttt{class\_field.spin\_reduce()}.}

In the \hpsrc\ code, we represent a spin matrix $W$ as a defined type,
\texttt{spinor}, which is encoded as a double array
\begin{equation}
(n; i_1,\ldots,i_n; w_1,\ldots,w_n) \equiv \sum_{k=1}^n w_k
\Gamma_{i_k}\;.
\end{equation}
It turns out to be significantly more efficient for the order of terms
in this array to not necessarily match a standard order of basis
elements $\{\Gamma_i\}$, hence the use of the index array $i_k$. In
particular, this allows us to omit basis elements with zero
coefficient.

Arithmetic operations have been overloaded to act appropriately on
objects of this type, including implemention of the multiplication
table. During inversion, an additional function argument,
\texttt{short\_spinor}, is used to employ the more efficient expression in
Eqn.~(\ref{eq_specialspinor_invert}).

\manual{\texttt{TYPE(spinor)} is implemented in the \hpsrc\ code in module
  \texttt{mod\_spinors.F90}. There is also a
  \texttt{TYPE(tayl\_spinor)} with \texttt{TYPE(taylor)}-valued
  amplitudes.}

\section{Even more realistic fermionic actions}
\label{sec_new_stuff}

Sec.~\ref{sec_old_stuff} summarised the general method that was
described in Ref.~\cite{Hart:2004bd}. In general, fermionic actions
are much more complicated than gluonic actions and several algorithmic
improvements are needed to efficiently calculate the associated
Feynman rules. We stress that by efficient we mean speed-ups of at
least an order of magnitude.

The algorithms described in this section can be used independently or
together, with the choice configurable at runtime of the code. All of
these features have also been implemented using taylor-valued
variables (as per Sec.~\ref{sec_taylor}) to provide automatic
differentiation of Feynman rules.

\subsection{Summands and factors}
\label{sec_summands_factors}

In many cases an action has a block-like structure that we can exploit
to make the evaluation of the reduced vertices more efficient. This is
particularly useful in the case of NRQCD and mNRQCD actions, which can
be heuristically written as $\bar{\psi}(1 - ABCA) \psi$. Such an
action can be expanded directly in the \hippy\ code but the extremely
large number of monomials makes this is inefficient (or impossibly
slow and memory-hungry). It is clear why: there is often little scope
for monomial compression between blocks. For instance, in (m)NRQCD,
blocks $AB$ and $CA$ live on different timeslices of the lattice, and
no compression is possible when combining them.

Instead, we recognise that the blocks are combined in a gauge
covariant manner, so that in $AB$, for instance, each contour in $B$
starts where a contour in $A$ finishes. Summing over the start/end
location of each block we obtain a convolution of terms and can use
the convolution theorem to construct the overall reduced vertex (i.e.
the momentum-space Fourier transform) from those of the individual
blocks.

We refer to the action as being a sum over terms that we call
``summands''. In the above example, there are two. Each summand is the
convolution of a number of ``factors'', with one factor in the first
summand and four in the second.

The overall reduced vertex $Y_{F,r}$ is the sum of the reduced
vertices for each summand. For each summand, $Y_{F,r}$ is calculated
by combining the reduced vertices $Y^{(k)}_{F,r}$ for each of $N$
factors that make up that summand, $k=1 \dots N$. We generate these by
expanding each factor of the action separately in the \hippy\ code,
with the convolution then carried out in the \hpsrc\ code.

Here we give the method for constructing $Y_{F,r}$ for a summand with
$N$ factors for general $r$. Expressions for specific $r=0 \dots 3$
are given in Appendix~\ref{sec_concrete_factors}.

In giving the general expression, we first establish some useful
notation. Consider the ordered set of the first $r$ integers:
$\{1,2,\dots,r\}$. We can form an ordered partition of this set: $\{
P_1,P_2,\dots,P_z \}$, where the cardinality $z \equiv \left| \{ ...
\} \right|$ is the number of elements in the partition. The set of all
such partitions we denote $P^{(r)}$. For instance, the partitions
$P^{(r)}$ for $r=1$,~$2$,~$3$ are shown in Table~\ref{tab_partitions}.
Note that we do not consider unordered partitions (e.g. $\{ \{1,3\},
\{2\} \}$) because the gauge fields are explicitly ordered in the
paths (Wilson lines) making up the action.

\begin{table}
\centering
\caption{\label{tab_partitions} The elements $P$ in the set of ordered
partitions, $P^{(r)}$, of the ordered set of the first $r$ integers,
$\{1,2,\dots,r\}$, for $r=1$,~$2$ and~$3$.}

\begin{tabular}[t]{ccc}
\hline
$r$ & $P$ & $|P|$ \\
\hline
1 & $\{ \{ 1 \} \}$ & $1$ \\
\hline
2 & $\{ \{ 1,2 \} \}$ & $1$ \\
  & $\{ \{ 1 \}, \{ 2 \} \}$ & $2$ \\
\hline
\end{tabular}
\hspace{5em}
\begin{tabular}[t]{ccc}
\hline
$r$ & $P$ & $|P|$ \\
\hline
3 & $\{ \{ 1,2,3 \} \}$ & $1$ \\ 
  & $\{ \{ 1,2 \},\{3 \} \}$ & $2$ \\
  & $\{ \{ 1 \}, \{ 2,3 \} \}$ & $2$ \\
  & $\{ \{ 1 \}, \{ 2 \}, \{ 3 \} \}$ & $3$ \\
\hline
\end{tabular}
\end{table}

The general reduced vertex for a summand with $N$ factors is then:
\begin{multline}
Y_{F,r}(\bld{p},\bld{q}; \bld{k}_1,\mu_1; \dots ;
\bld{k}_r,\mu_r) =
\\
\sum_{P \in P^{(r)}} \; 
\sum_{1 \le n_1 < n_2 < \dots < n_{|P|} \le N} \;
\left\{
\prod_{Q \in P} \;
\left[
\left(
\prod_{k = n_{i-1}+1}^{n_i-1}
Y_{F,0}^{(k)}(\bld{p}_i,-\bld{p}_i)
\right)
\right.
\right.
\\
\left.
\times
Y_{F,|Q|}^{(n_i)}(\bld{p}_i,-\bld{p}_{i+1};
\bld{k}_{Q_1},\mu_{Q_1}; \dots ; \bld{k}_{Q_{|Q|}},\mu_{Q_{|Q|}})
\right]
\\
\left.
\times
\left(
\prod_{k = n_{|P|}+1}^{N}
Y_{F,0}^{(k)}(-\bld{q},\bld{q})
\right)
\right\}
\label{eqn_factors}
\end{multline}
where $i=1 \dots |P|$ is the position of element $Q$ in the ordered
set $P$ and $n_0 = 0$. Momentum is conserved in the diagram, so
$\bld{p}_1 = \bld{p}$ and subsequent $\bld{p}_{i+1} =
\bld{p}_i + \bld{k}_Q$ (with $i$ defined as above). Here
$\bld{k}_{Q}$ refers to the summed momenta for the set $Q$ (e.g.
$\bld{k}_{\{1,2\}} \equiv \bld{k}_1 + \bld{k}_2$), implying
$\bld{p}_{|P|} = -\bld{q}$.

\manual{In the \hpsrc\ code, the $Y_{F,r}$ for each summand are
calculated and combined in \texttt{SUBROUTINE vertex\_qq*()}, with the
reduced vertex for each factor calculated in \texttt{SUBROUTINE
vertex\_qq*\_partial()}.}

\subsection{Two-level actions}
\label{sec_two_level}

Fermion actions often use fattened links to reduce discretisation
errors in numerical simulations. By fattening or smearing a link, we
will, in general, end up with an $N \times N$ complex colour matrix
$M$, which can be expanded in powers of the gauge field $A_\mu$. It is
often the case that this matrix is reunitarised by projecting it back
onto the gauge group $SU(N)$ or, more usually, simply back onto the
related group $U(N)$. Fattened links can then be further fattened, in an
iterative procedure. An example of this is the HISQ action.

To complement these numerical simulations, we need to do perturbative
calculations using the same actions. We confine our attention here to
the HISQ action (and simpler variants of the same form, for testing),
with an iterated, two-level smearing procedure with an intermediate
reunitarisation:
\begin{equation}
U^\textrm{HISQ} = ( F_{\textrm{ASQ}'} 
\circ P_{U(3)} \circ F_{\textrm{FAT7}} )[U]
\end{equation}
where $U=\exp(gA)$ is the unsmeared gauge field, $P_{U(3)}$ denotes
the polar projection onto $U(3)$ (as used in simulations, and
\textit{not} $SU(3)$
\cite{Bazavov:2009jc}),
and the FAT7 and ASQ' (a slightly modified version of ASQ) smearings
are defined in
Ref.~\cite{Follana:2006rc}.

Straightforward application of the expansion algorithm above is
theoretically possible but practically unfeasible: the number of
monomials is huge and the memory requirements of the \hippy\ code
quickly become excessive. We get around this by taking advantage of
the two-level structure inherent in the definition of the action and
that the intermediate reunitarisation. This allows us to express the
partially-smeared gauge field as a member of the associated gauge Lie
algebra, allowing us to split the derivation and subsequent
application of the Feynman rules into two steps.

In the first step, the Feynman rules for the outer (or ``top'') layer,
the ASQ' action, are derived in the same way as before. Representing
the reunitarised (and so far uncalculated) FAT7R smeared links by a
new, Lie-algebra--valued gauge potential, $B_\mu$:
\begin{equation}
U_\mu^{\textrm{FAT7R}}(x) = ( P_{U(3)} \circ F_{\textrm{FAT7}}
)[U_\mu] = e^{B_\mu(x+\tfrac{1}{2}\hat\mu)} \;, 
\end{equation}
we can use the \hippy\ code to expand the ASQ' action in terms of
$B_\mu$ as before. We obtain similar position-space contributions
\begin{equation}\label{eqn:asqfr}
V_r=\frac{g^r}{r!}\sum_i f^{\textrm{ASQ'}}_{r;i} 
\bar{\psi}(x_{r;i})B_{\mu_1}(v_{r;i,1})\cdots B_{\mu_r}(v_{r;i,r})
\Gamma_{r;i}\psi(y_{r;i})
\end{equation}
that Fourier transform to give monomials of the usual form.

To complete the derivation of the HISQ Feynman rules, we also need to
know the expansion of $B_\mu$ in terms of the original gauge potential
$A_\mu$. To obtain this, we write inner (or ``bottom'') layer, the
FAT7-smeared link, as $F_{\textrm{FAT7}}[U] = M = HW$, where
$H^\dag=H$ and $W \in U(3)$ (we suppress Lorentz and lattice site
indices in the following). We again use the \hippy\ code to obtain an
expansion
\begin{equation}
M = c[{\bf 1} + a_\mu A_\mu + a_{\mu\nu} A_\mu A_\nu + \ldots]
\label{eqn_M_reunit}
\end{equation}
where $c$ is a normalisation constant which, for simplicity in the
text, we assume has been rescaled to unity. The notation here is, for
instance,
\begin{equation}
a_{\mu\nu} A_\mu A_\nu = \sum_{x,y}
a_{\mu\nu}(x,y)A_\mu(x+\frac{1}{2}\hat\mu)
A_\nu(y+\frac{1}{2}\hat\nu) \; .
\end{equation}
Then unitarity of $W$ implies that $R\equiv MM^{\dag}=H^2$ and hence
$W=R^{-1/2}M$ using the expansion
\begin{equation}
R^{-1/2}=(1+(R-1))^{-1/2}=1-\frac{1}{2}(R-1)+\frac{3}{8}(R-1)^2+\ldots
\end{equation}
Rearranging the result as $W=\exp(B)$, i.e.
\begin{equation}
B=\log(W)=(W-1)-\frac{1}{2}(W-1)^2+\ldots
\label{eqn_algebra2}
\end{equation}
finally yields the desired expansion of $B$. These formul\ae\ are
implemented in this form in the \hippy\ code.

Given this, we can now numerically reconstruct the HISQ Feynman rules
for any given set of momenta from Eqn.~(\ref{eqn:asqfr}) by a convolution
of the ASQ' Feynman rules of Eqn.~(\ref{eqn:asqfr}) with the expansion of
$B_\mu$ in terms of $A_\mu$,  summing up all the different ways in which
the gluons $A_\mu$ going into the vertex could have come from the fields
$B_\mu$ appearing in Eqn.~(\ref{eqn:asqfr}).

In more detail, we now separately have the expansions of the ASQ'
action in terms of the fattened gauge fields $B$, and of $B$ in terms
of the unfattened gauge field $A$. To obtain the correct reduced
vertex, we must find all the ways that we can get unfattened gluons of
the correct Lorentz polarisations (``directions''). In doing this, we
bear in mind that $B_\mu$ contains, in principle, gauge fields $A_\nu$
in all directions and not just $\nu = \mu$. Below we give an
expression for the reduced vertex for general $r$. Explicit formul\ae\
for $r \le 3$, as implemented in the \hpsrc\ code, are given in
Appendix~\ref{sec_concrete_algebra}. Using the partitions $P^{(r)}$ as
before, the reduced vertex for a two-level action is::
\begin{multline}
Y_{F,r}(\bld{p},\bld{q}; \bld{k}_1,\mu_1; \dots ;
\bld{k}_r,\mu_r) = 
\\
\sum_{P \in P^{(r)}} \;
\sum_{\nu_1, \dots, \nu_{|P|}} \;
Z_{F,|P|}(\bld{p},\bld{q}; \bld{k}_{P_1},\nu_1; \dots ;
\bld{k}_{P_{|P|}}, \nu_{|P|})
\\
\times \prod_{Q \in P} \;
X_{F,|Q|}^{\nu_{i}}    (\bld{k}_{Q_1},\mu_{Q_1}; \dots ; 
\bld{k}_{Q_{|Q|}},\mu_{Q_{|Q|}} )
\label{eqn_algebra}
\end{multline}
where $i=1 \dots |P|$ is again the position of element $Q$ in the
ordered set $P$. In the algebra reduced vertex $X$, the momenta
$\bld{k}_{Q_i}$ are each one of the arguments to the overall
reduced vertex $Y$. As before, in the field reduced vertex $Z$,
$\bld{k}_{P_j}$ refers to the summed momenta for the partition
$P_j$.

\manual{In the \hpsrc\ code, the $Y_{F,r}$ are calculated in
\texttt{SUBROUTINE vertex\_qq*\_partial()}.}

\begin{figure}
\centering
\includegraphics[height=0.9\textwidth,clip,angle=-90]{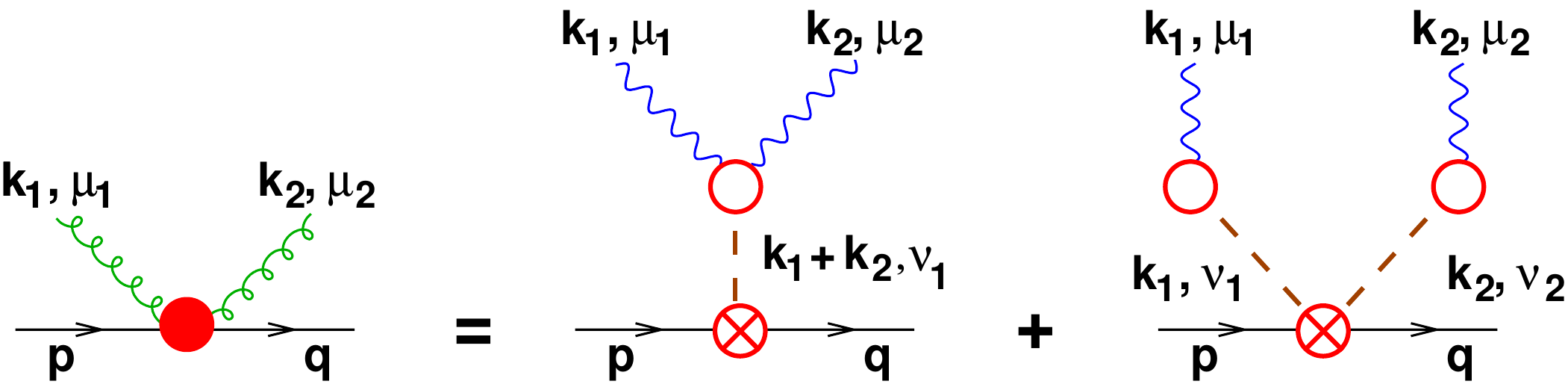}
\caption{\label{fig_upper_lower}A graphical representation of the 
reduced vertex $Y_{F,r}$ for a two-level action, with $r=2$ as in
Eqn.~(\ref{eqn_Y_F2}). Solid circles represent the $Y_{F,r}$, whilst
crossed and open circles represent $Z_{F,r}$ and $X_{F,r}$
respectively. Top-level, fattened gluons $B_\mu$ are represented by
dashed (brown) lines, whilst bottom-level, unfattened gluons $A_\mu$
are shown as wavy (blue) lines. The two sub-diagrams represent the two
partition contributions listed in Eqn.~(\ref{eqn_Y_F2}). All momenta
are incoming to the vertex.}
\end{figure}

In Fig.~\ref{fig_upper_lower} we represent this expansion graphically
for $r=2$. Solid circles represent the $Y_{F,r}$, whilst crossed and
open circles represent $Z_{F,r}$ and $X_{F,r}$ respectively.
Top-level, fattened gluons $B_\mu$ are represented by dashed (brown)
lines, whilst bottom-level, unfattened gluons $A_\mu$ are shown as
wavy (blue) lines. The two sub-diagrams represent the two partition
contributions listed in Eqn.~(\ref{eqn_Y_F2}).

As we shall discuss later, this partitioning translates naturally into
blocks of code. For certain calculations, symmetries of the Feynman
diagram will lead to the contributions of some of these blocks being
zero. We can then improve the performance of the code by commenting
them out in these circumstances. For instance, in the ``tadpole''
diagram in the one-loop fermion self energy, we can remove the term in
$Y_{F,2}$ that is proportional to $X^{\nu_1}_{F,2}$
\cite{Lee:2002fj}.
Similarly, in the calculation of the radiative corrections to the
gauge action
\cite{Hart:2008zi},
we can suppress the term proportional to $X^{\nu_1}_{F,3}$ in the
``octopus'' diagram.

We can further optimise the calculation of the reduced vertices
$Y_{F,r}$ by reusing terms $X^\nu_{F,r}$ that appear multiple times in
the expressions. For instance, in Eqn.~(\ref{eqn_Y_F3}) in
Appendix~\ref{sec_concrete_algebra} we can reuse
$X_{F,1}^\nu(\bld{k}_1,\mu_1)$ and $X_{F,1}^\nu(\bld{k}_3,\mu_3)$.

For testing it is useful to define a simpler variant of the smearings
described on page~3 of
Ref.~\cite{Orginos:1999cr},
which we call ``FAT3''. It is composed just of a central link and
adjacent 3-staples with weights $c_1 = \frac{1}{2}$, $c_3 =
\frac{1}{12}$.

\subsubsection{$Z_{F,r}$: the field reduced vertex}

The field reduced vertex essentially calculates the Feynman rule for
obtaining $r$ \textit{fattened} gluons in particular directions
$\mu_1, \dots , \mu_r$. The effect of fattening is treated separately
in the $X_{F,r}$.

$Z_{F,r}$ is composed of a sum of $n_r$ monomials, each representing a
different way of obtaining the required gluons from the contours
comprising the action:
\begin{multline}
Z_{F,r}(\bld{p},\bld{q}; \bld{k}_1,\mu_1; \dots ;
\bld{k}_r,\mu_r) = 
\\
\frac{1}{r!}
\sum_{n=1}^{n_r} \Gamma_n f_n \exp \left( \frac{1}{2} \left(
\bld{p} \cdot \bld{x} + 
\bld{q} \cdot \bld{y} + \sum_{j=1}^r
\bld{k}_j \cdot \bld{v}_{n,j} \right) \right)
\label{eqn_zy}
\end{multline}
where $\Gamma_n$ is a spin matrix.

Note that if there is no fattening of the underlying gluons, this is
precisely the reduced vertex $Y_{F,r}$ described in
Eqn.~(\ref{eqn_sy}), if we incorporate the $1/r!$ factor that was in
Eqn.~(\ref{eqn_SS}).

\manual{In the \hpsrc\ code, the $Z_{F,r}$ are calculated in
\texttt{SUBROUTINE vertex\_qq*\_partial\_noalg()}.}

\subsubsection{$X_{F,r}$: the algebra reduced vertex}

If the gluons are fattened, the effect of this is contained in the
reduced vertices $X_{F,r}$. These represent how we would choose $r$
unfattened gluons in directions $\mu_1,\dots\mu_r$ from a fattened
link in the $\nu$ direction:
%
\begin{equation}
X_{F,r}^{\nu} (\bld{k}_1,\mu_1; \dots ;
\bld{k}_r,\mu_r) = 
\\
\frac{1}{r!}
\sum_{n=1}^{n_r} f_n \exp \left( \frac{1}{2} \sum_{j=1}^r
\bld{k}_j \cdot \bld{v}_{n,j}^\prime \right)
\label{eqn_xy}
\end{equation}
%
Note that there is no spin matrix in $X$, so the order of the
multiplication in the expression for $Y$ does not matter.

The position vectors $\bld{v}_{n,i}^\prime \equiv \bld{v}_{n,i}
- \bld{e}_{\nu}$ refer to the position of the unfattened link
relative to the start of the fattened. As such, we need to remove the
half-link shift that was implicit in the definition of
$\bld{v}_{n,i}$ as the \textit{centre} of the link in the basic
expansion algorithm.

The start and end sites of the contour $\bld{x}$ and $\bld{y}$
do not appear in $X$, making $X^\nu_{F,r=0}=1$ for $r=0$.

Of course, the actual values of $n_r$, $f_n$ and $\{v_{n,r}\}$ will be
different in Eqns.~(\ref{eqn_zy}) and~(\ref{eqn_xy}), as they come
from different smearing prescriptions.

\manual{In the \hpsrc\ code, the $X_{F,r}$ are calculated in
\texttt{SUBROUTINE vertex\_qqg*\_algebra()}. $X_{F,r=0}$ is not
explicitly calculated. The subtraction $\bld{v}_{n,i}^\prime =
\bld{v}_{n,i} - \bld{e}_{\mu_i}$ is done in the \hippy\ code
before writing the vertex files to disk.}

\subsection{Hardwired reunitarisation}
\label{sec_reunit}

We described above how splitting a fermion action such as HISQ into
two parts simplifies the generation of the Feynman rules. This allowed
us to exploit the reunitarisation of the inner FAT7R smeared links
allows them to be expressed in terms of a new gauge potential $B$
(suppressing the Lorentz index). 

Nonetheless, the Feynman rules for $B$ (as calculated in $X_{F,r}$)
are still too complicated to derive for $r \ge 3$. The reason is clear
if we look a tthe formula for the reunitarisation. Using
Eqn.~(\ref{eqn_M_reunit}), the reunitarised field is $W = \left( M
M^\dagger
\right)^{-\frac{1}{2}} M$, which we can express as an
element of the algebra $B = \log W$:
\begin{align}
B & \equiv b_\mu A_\mu + b_{\mu \nu} A_\mu A_\nu + b_{\mu \nu \sigma}
A_\mu A_\nu A_\sigma + \dots
\nonumber \\
b_\mu & = a_\mu
\nonumber \\
b_{\mu \nu} & = \frac{1}{2}\left( a_{\mu \nu} - a^\dagger_{\mu \nu} \right) 
\nonumber \\
b_{\mu \nu \sigma} & = 
\frac{1}{2}\left( a_{\mu \nu \sigma} - a^\dagger_{\mu \nu \sigma} \right)
-\frac{1}{4} \left(a_\mu \left[ a_{\nu \sigma} + a^\dagger_{\nu \sigma} \right]
+ \left[a_{\mu \nu} + a^\dagger_{\mu \nu}\right] a_\sigma \right) 
+\frac{1}{3} a_\mu a_\nu a_\sigma 
\label{eqn_reunit}
\end{align}
If there are $n$ monomials in $a_\mu$, there will be at least $n^3$ in
$b_{\mu \nu \sigma}$ arising from the term $a_\mu a_\nu a_\sigma$.
There will be some compression, but this will at best only reduce this
number by a factor of around~2. For the simpler smearing FAT3, $n=19$
and already the \hippy\ code struggles to produce $b_{\mu
\nu \sigma}$ for FAT3R. For FAT7, $n=135$ and direct reunitarisation 
to produce FAT7R links using the \hippy\ code is out of the question.

\manual{In the \hippy\ code, the reunitarisation by projection to 
$U(N)$ is done in \texttt{proj\_SU3()} in \texttt{class\_field.py}.}

The alternative is to use the \hippy\ code to produce the $\{a\}$ and
to do the reunitarisation in the \hpsrc\ code by hardwiring the
formul\ae\ in Eqn.~(\ref{eqn_reunit}). As before, the $X_{F,r}$
algebra reduced vertices are based on expansion coefficients $\{b\}$.
The inputs from the \hippy\ code, however, implement the coefficients
$\{a\}$. We use these to build a set of algebra reduced vertices
$W_{F,r}$ and Eqn.~(\ref{eqn_reunit}) gives the relation between the
reunitarised $X_{F,r}$ and the $W_{F,r}$:
\begin{gather}
X^\nu_{F,1}(\bld{k}_1,\mu_1) = W^\nu_{F,1}(\bld{k}_1,\mu_1)
\nonumber \\
X^\nu_{F,2}(\bld{k}_1,\mu_1;\bld{k}_2,\mu_2) =
\frac{1}{2} \left(
W^\nu_{F,2}(\bld{k}_1,\mu_1;\bld{k}_2,\mu_2) - 
W^\nu_{F,2}(\bld{k}_2,\mu_2;\bld{k}_1,\mu_1) \right)
\nonumber \\
X^\nu_{F,3}(\bld{k}_1,\mu_1;\bld{k}_2,\mu_2;\bld{k}_3,\mu_3) =
\nonumber \\
\begin{split}
\frac{1}{2} \left(
W^\nu_{F,3}(\bld{k}_1,\mu_1;\bld{k}_2,\mu_2;\bld{k}_3,\mu_3) + 
W^\nu_{F,3}(\bld{k}_3,\mu_3;\bld{k}_2,\mu_2;\bld{k}_1,\mu_1)
\right)
\\
-\frac{1}{4} \left( W^\nu_{F,1}(\bld{k}_1,\mu_1) \left[
W^\nu_{F,2}(\bld{k}_2,\mu_2;\bld{k}_3,\mu_3) +
W^\nu_{F,2}(\bld{k}_3,\mu_3;\bld{k}_2,\mu_2) \right]
\right.
\\
\left.
 + \left[
W^\nu_{F,2}(\bld{k}_1,\mu_1;\bld{k}_2,\mu_2) +
W^\nu_{F,2}(\bld{k}_2,\mu_2;\bld{k}_1,\mu_1) \right] 
W^\nu_{F,1}(\bld{k}_3,\mu_3) \right)
\\
+\frac{1}{3} W^\nu_{F,1}(\bld{k}_1,\mu_1)
W^\nu_{F,1}(\bld{k}_2,\mu_2)
W^\nu_{F,1}(\bld{k}_3,\mu_3) \; .
\end{split}
\label{eqn_wy}
\end{gather}

\manual{In the \hpsrc\ code, the $W_{F,r}$ are calculated in
\texttt{SUBROUTINE y\_qqg*\_alg\_basic()}.}

\section{Description of the software}
\label{sec_software}

\subsection{The \hippy\ code}

The \hippy\ code is used to Taylor expand lattice actions, producing
output files encoding the monomials making up the reduced vertices
that can later be used by the \hpsrc\ code to evaluate Feynman
diagrams and integrals.

The code is written in Python, for which interpreters are freely
available for a wide range of computational platforms
\cite{ref_python}.

\subsubsection{Installation and compatibility}

The \hippy\ code can be run on any machine which has Python version
2.5.x installed on it. It is also expected to work with any version
2.x, but is not (yet) compatible with version 3.x. Only the packages
supplied in a standard installation of Python are required.

Installation consists of unpacking the source files in the chosen
location. 

\subsubsection{Testing}

A set of unit test programs are contained in subdirectory
\texttt{tests}. They are run from the command line with no options and
the code is expected to pass all of these. No input files are needed
for the tests.

Whilst the unit testing does not cover each code feature separately,
those in \verb|test_class_field.py| do cover most features combined.

\subsubsection{Main programs}

The \hippy\ code consists of the following main programs that make use
of the core routines described below.

A full list of options for each of the main programs can be obtained
by running them from the command line with option \texttt{--help}.

\paragraph{\texttt{sample\_link.py}}

Used to generate monomials from the algebra of the group
for a single fattened link in the specified direction. These are used
to evaluate $X_{F,r}$ in Eqn.~(\ref{eqn_xy}) or $W_{F,r}$ in
Eqn.~(\ref{eqn_wy}) in the \hpsrc\ code. 

The fattening style is specified using \texttt{--bottom\_style}. The
output filenames have the form \texttt{algebra\_\%s\%i\_qq*.in}, where
\texttt{\%s} is replaced by the name of the fattening style and
\texttt{\%i} by the direction of the link (with value
\verb|parameters.D| mapped to zero). Wildcard \texttt{*} is $r$ copies
of the character ``\texttt{g}'' (up to maximum \texttt{parameters.R}).

For monomials that will be used in Eqn.~(\ref{eqn_wy}) (with hardwired
reunitarisation carried out in the \hpsrc\ code), the option
\texttt{--fld\_as\_alg} should be specified. In this case \texttt{\%s}
is a concatenation of character ``\texttt{F}'' and the name of the
fattening.

If an undefined fattening style is specified with
\verb|--bottom_style|, a list of acceptable styles is printed.

\paragraph{\texttt{sample\_naive.py}}

Used to generate monomials for the naive (or, equivalently,
staggered) lattice Dirac action using fattened links. These are used
to evaluate $Y^{(k)}_{F,r}$ in Eqn.~(\ref{eqn_factors}) in the \hpsrc\
code.

The fattening style(s) are specified using \texttt{--top\_style} and
\texttt{--bottom\_style}, and should not include fattening already
carried out at the algebra stage above. The output filenames have the
form \texttt{vertex\_\%s\_qq*.in}, where \texttt{\%s} is replaced by
the name of the top fattening style and \texttt{*} is $r$ copies of
the character ``\texttt{g}''. For complicated fattening prescriptions,
there is likely to be some filename clashes, so care should be
exercised.

If an undefined fattening style is specified with
\verb|--top_style|, a list of acceptable styles is printed.

\paragraph{\texttt{sample\_glue.py}}

Used to generate monomials for a range of simple gauge actions,
specified by a command line argument. As described above, these
vertices have been (partially) symmetrised.

\subsubsection{Example of use}

As an example, here are the commands one would use to generate the
four-dimensional HISQ reduced vertex files using a two-level action
with hardwired reunitarisation and mass $am = 0.2$:

\begin{verbatim}
python sample_link.py --bottom_style fat7 --fld_as_alg 1
python sample_link.py --bottom_style fat7 --fld_as_alg 2
python sample_link.py --bottom_style fat7 --fld_as_alg 3
python sample_link.py --bottom_style fat7 --fld_as_alg 4
python sample_naive.py --top_style asq_for_hisq 0.2
\end{verbatim}

\noindent
which creates files \texttt{algebra\_Ffat7\%i\_qq*.in} and
\texttt{vertex\_asq\_for\_hisq\_qq*.in}. The same Feynman
rules can be generated using

\begin{verbatim}
python sample_naive.py --top_style asq_for_hisq \
   --bottom_style fat7r 0.2
\end{verbatim}

\noindent
but this generates \textit{far} more monomials, which soon becomes
prohibitive as $r$ increases, as discussed above.

For HISQ fermions, there is the option to set the mass--dependent
parameter $\varepsilon$ to zero for small $am$. This is done using a
command line option \verb|--hisq_eps_zero| for \verb|sample_naive.py|.

For main programs to calculate Feynman rules for other actions, e.g.
NRQCD, interested readers should contact the authors.

\subsubsection{Core routines}

These main programs make use of the following core routines:

\paragraph{\texttt{parameters.py}}

Defines physical and numerical parameters associated with the
expansion. If the monomials require complex-valued amplitudes, e.g.
for mNRQCD actions, this is enabled here.

\paragraph{\texttt{template.py}} 

Defines some link fattening prescriptions that can be used to define
actions. Each fattening is a set of paths which do not need to be
gauge covariant, so three-link Naik terms can be included, for
instance. Instructions for adding new fattenings are included in this
file.

\paragraph{\texttt{spin\_multiplication.py}}

Implements the implicit representations of the spin algebras described
in Sec.~\ref{sec_spinor}. Running this from the command line displays
the multiplication table from Eqn.~(\ref{eqn_spin_mult}) for either
Pauli or Dirac spin matrices.

\paragraph{\texttt{wilson.py}}

The main expansion engine that converts a collection of
Wilson lines into a Taylor expansion in the form of reduced vertices
built from monomials. The collection of Wilson lines is defined by

\begin{verbatim}
thepath = [ [c1,c2,\dots],[p1,p2,\dots] ]
\end{verbatim}

\noindent
where \texttt{c1}, \texttt{c2} etc. are the amplitude of each contour,
the path of which is defined by a set of signed integers. For
instance, a plaquette would be \texttt{p1=[1,2,-1,-2]}.

As an example of a complete path, a two-dimensional naive fermion
action would be defined by a path:

\begin{verbatim}
thepath = [ [0.5,-0.5,0.5,-0.5,m],[ [1],[-1],[2],[-2],[] ] ]
\end{verbatim}

\noindent
The expansion is not symbolic, so the mass \texttt{m} must take a
specific numerical value. 

It is assumed that all links are of the same, fattened style. The
fattening can contain any number of iterated styles defined in
\texttt{template.py}, given as a list \texttt{top\_style} interpreted
from right to left.

At the lowest level of this iteration, a further single level of
fattening can be specified using \texttt{bottom\_style}. Uniquely, a
reunitarisation by projection onto $U(N)$ can be imposed after this
fattening by appending character ``r'' to the end of the name of the
style (again, chosen from the definitions in \texttt{template.py}).

\paragraph{\texttt{class\_entity.py}}

Each monomial in the reduced vertex is encoded as an instance of this
class, a description of which can be found in Sec.~4 of
Ref.~\cite{Hart:2004bd}.
Translation invariance is exploited so that all paths are assumed to
start at the origin, $\bld{x} = \bld{0}$. If this is not the case (e.g.
for open boundary conditions in the temporal direction for
Schr\"{o}dinger functional calculations
\cite{Takeda:2007dt,Takeda:2008rr}),
then an attribute \texttt{start\_site} would need to be introduced to
this class and appropriate changes made to ensure gauge covariance in
the multiplication and in other places.

\paragraph{\texttt{class\_field.py}}

The collection of monomials defining the reduced vertices returned by
\texttt{wilson.path()} are stored as a dictionary in an instance of
this class. A fuller description can be found in Sec.~4 of
Ref.~\cite{Hart:2004bd}.

\paragraph{\texttt{class\_algebra.py}}

Very similar in form to \texttt{class\_field.py}, this class contains
the reduced vertices associated with the algebra in
Eqn.~(\ref{eqn_algebra2}).

\paragraph{\texttt{mathfns.py}}

contains a collection of methods not tied to any particular class.

\subsection{The \hpsrc\ code}

The \hpsrc\ code uses the input files from the \hippy\ code to create
Feynman rules, and combines these to evaluate Feynman integrals by
exact mode summation or statistical estimation using the
\vegas\ algorithm
\cite{Lepage:1977sw,Lepage:1980dq}.

\subsubsection{Installation and compatibility}

The \hpsrc\ code requires a standards-compliant Fortran95 compiler
with support for minimal CPP directives (chiefly \texttt{\#ifdef} and
\texttt{\#ifndef}). The command \texttt{make} is optional but useful.
An appropriate MPI wrapper for the compiler will be needed for
creating a parallel version of the executable. No additional libraries
are required.

The code is known to compile and execute correctly with the following
compilers: Intel ifort, Portland pgf90, NAG f95 and GNU gfortran on
Unix-based systems (the first two also with MPI on parallel machines)
and Silverfrost/Salford ftn95 and Lahey-Fujitsu lf95 on Microsoft
Windows systems using cygwin.

To install the code, the source should be unpacked in a suitable
directory and the file \texttt{Makefile\_details} (which contains
machine-dependent information) edited to point to the correct
compiler.

It is a good idea to execute \texttt{make clean} before compiling the
code. A given target \texttt{this\_executable} can be compiled using
the command 
\begin{verbatim}
make this_executable
\end{verbatim} 
\noindent
Use the command
\texttt{make} to see a list of possible executables, or examine
\texttt{Makefile}.

\subsubsection{\hpsrc\ filename conventions}

Main programs are named \texttt{main\_*.F90} and compile to form
executables with the same filestem and extension \texttt{.\$(EXE)} as
specified by \texttt{Makefile\_details}. Typically \texttt{\$(EXE) =
x}, but certain compilers expect \texttt{\$(EXE) = exe}.

A Fortran module \texttt{this} is typically contained in a file
named \texttt{mod\_this.F90}.

Input files containing information that should be specified at compile
time are named with extension \texttt{.i}. The command \texttt{make
clean} must be executed before recompiling if any of these files are
changed. These files typically contain unavoidable CPP directives and
the number of these has been kept to a minimum. 

Input files that are read at runtime are have names with extension
\texttt{.in}. Most input files have a filename that reflects that of
the module that reads them.

\subsubsection{Testing}

The code has been rigorously tested as a whole in a number of
scientific calculations, being compared with identical calculations
carried out using different programs.

In addition, various manual and automatic tests are routinely carried
out to verify the continued correctness of the code. The manual tests
are located in subdirectory \verb|tests/|.

\paragraph{Automatic differentiation}

Without a templating option in Fortran, each vertex is effectively
coded twice: once as a complex-valued object and once as a
taylor-valued object that includes the automatic derivatives.

The code exploits this by performing a finite difference of the first
to compare with the derivatives included in the second. To avoid
multiplication of errors in higher-order difference operators, the
tests are carried out as follows. First, the non-taylor version is
compared with the zero-order derivative from the taylor version. If
this agrees, a first-order, symmetric difference is made of the
zero-order derivative in the taylor in direction $\tau$. We check
that, for all $\tau$, this agrees (within a specified tolerance) with
the analytic first order derivative in the taylor object. If so, we
form first-order differences of the analytic first-order derivatives
and compare these with the analytic second order derivatives and so
on.

These comparisons are done at runtime when the vertex modules are
first initialised.

This test checks not only that the automatic differentiation (and
associated overloading of operators) is working correctly, but also
that there are no coding inconsistencies between the taylor and
non-taylor versions of the vertices. It cannot, of course, detect
algorithmic errors that have been consistently coded in both versions.
An additional, lower level test of the TaylUR package is provided by
\verb|test_taylors.x|.

\paragraph{\texttt{test\_compare\_vertices.x}}

Can be used to check that the Feynman rules are the same for two
different implementations of the same quark action. This provides a
rigorous check of the algorithms in \verb|mod_vertex_qq*.F90|.
Having compiled \verb|test_compare_vertices.x|, a Python script
\verb|test_compare_vertices.py| can be executed to complete various
independent checks of the two--level action decomposition, the
hardwired reunitarisation and the summand/factor division of the
action. The last is based on an NRQCD action and also provides a
strong test of the implementation of the Pauli spin algebra.

\paragraph{Tests of \vegas}

The \vegas\ code is automatically tested at runtime, evaluating the
area under a two-dimensional, normalised Gaussian as well as a contour
integral. These tests can also be run using \verb|test_vegas.x|.

\paragraph{Lower level \hpsrc\ tests}

\verb|test_spinors.x| tests type \verb|spinor| defined in
Sec.~\ref{sec_spinor}. In addition, \verb|test_pauli_dirac_spinors.x|
tests the embedding of Pauli two-spinors into Dirac 4-spinors.
\verb|test_print_vertices.x| prints the values of the fermionic
vertices for a random set of momenta. The same random number seed is
used each time, so running the code for different
\verb|vertex_qq_composite.in| on the same machine will yield outputs
that can be directly compared.

\subsubsection{Main programs}

To illustrate the use of the \hpsrc\ code, we provide three sample
programs with Makefile targets:

\paragraph{\texttt{quark\_sigma.x}}

Calculates the one-loop renormalisation of ASQTAD improved staggered
fermions. To do this, it evaluates the self-energy Feynman diagrams
and their derivatives.

\paragraph{\texttt{nrqcd\_sigma.x}}

Performs a similar calculation for heavy NRQCD fermions.

\paragraph{\texttt{nflwimp.x}}

Calculates the HISQ fermion contribution to the one-loop radiative
corrections to the L\"{u}scher-Weisz gauge action, as per
Refs.~\cite{Hart:2008zi,Hart:2008sq}.

\subsubsection{Core routines}

This subsection describes modules that are needed to use the vertex
modules. If MPI is to be used, the file \verb|use_mpi.i| must be
changed prior to compilation. Similarly, if monomials require complex
amplitudes, e.g. for mNRQCD calculations, this is enabled in
\verb|use_complex_amplitudes.i|.

\paragraph{\texttt{mod\_num\_parm.F90}}

Numerical parameters that are used by most of the other
modules. Most parameters in this module will not need to be
changed. It also optionally reads from file
\texttt{paths.in} the pathnames of the directories holding the vertex
and algebra files generated by the \hippy\ code.

\paragraph{\texttt{mod\_phys\_parm.F90}}

Physical parameters used in the calculation. The code
has only been tested for number of dimensions \texttt{Ndir} $= 4$ and
colours \texttt{Ncol} $= 3$. Other choices will require careful
testing before use.  Values for many parameters can be changed at
runtime by editing file
\texttt{phys\_parm.in}
Twisted boundary conditions are selected by setting the number of
twisted directions \texttt{ntwisted\_dirs} to 2, 3 or 4. Value 0 gives
periodic boundary conditions.
For finite lattices, the lattice size is specified in
\texttt{latt\_size(0:Ndir-1)}, with the first component the temporal
one. If momenta are to be squashed $k_\mu \to k_\mu - \alpha_\mu \sin
(k_\mu)$
\cite{Luscher:1985wf}, 
the squash factors $\alpha_\mu$ in each direction are specified in
\texttt{mom\_squash(0:Ndir-1)}.
The anisotropy metric \texttt{anis} is only
supported currently for gluonic vertices.

\paragraph{\texttt{mod\_momenta.F90}}

Defines type \texttt{mom} which encodes momenta. Each instance
contains the momentum components, the twist vector (only relevant for
twisted boundary conditions) and a route variable that is used in
automatic differentiation.

\paragraph{\texttt{mod\_taylors.F90}}

Defines type \texttt{taylor}, as described in
Sec.~\ref{sec_taylor}.

\paragraph{\texttt{mod\_spinors.F90}}

Defines types \texttt{spinor} and \texttt{tayl\_spinor}, as described
in \\
Sec.~\ref{sec_spinor}.

\paragraph{\texttt{mod\_matrices.F90}}

Defines types \texttt{mat4x4}, used to explicitly represent objects
such as the gluon propagator in Lorentz index space. A corresponding
taylor-valued type \verb|tayl_mat4x4| is also defined.

\paragraph{\texttt{mod\_colour.F90}}

Calculates the Clebsch-Gordan factors for both gluonic and fermionic
vertices, for periodic and twisted boundary conditions.
Colour factors for periodic boundary condition are pre-computed and
stored in the compile-time files \texttt{tr\_cols\_N.i} and
\texttt{mat\_cols\_N.i}. For twisted boundary conditions, the
Clebsch-Gordan factors are computed as needed.

\paragraph{\texttt{mod\_mod\_mpi.F90}}

The code can either be run as a scalar code on a single processor or
as a parallel code using MPI. To achieve this with the minimum of
compile-time code modifications (either by hand or using CPP
directives), \texttt{mod\_mod\_mpi.F90} contains a set of dummy MPI
subroutines. Whether these dummy routines are used or not is
controlled by the single CPP directive in file \texttt{use\_mpi.i}.
Note that \texttt{mod\_mod\_mpi.F90} is the only file that reads
\texttt{use\_mpi.i}.
Only the MPI commands we use are included in \texttt{mod\_mod\_mpi.F90}.
We do not know of any externally maintained library of dummy MPI
commands for use in such situations. If this were to exist, it could
easily be used to replace this file.

Throughout the code we are (or try to be) careful that input/output
files are only opened by one processor, with information then shared
using MPI commands.

\paragraph{\texttt{mod\_vegasrun.F90}}

Provides an implementation of the \vegas\ algorithm, based on the
original code of Peter Lepage 
\cite{Lepage:1977sw,Lepage:1980dq},
who also helped convert the code to a parallel version using MPI. On
parallel machines, the processors are divided into farms, each of
which does an independent \vegas\ estimate of the integral. Within the
farm, all processors carry out the same calculation, interacting only
to share the work of evaluating the integrand at each of the Monte
Carlo points. The number of processors per farm is controlled in file
\verb|vegas_parm.in|, and must be a factor of the number of
processors. Farming has the advantage of avoiding potentially slow
global communications on large clusters.

An improved parallel implementation is currently being prepared as
part of a DEISA DECI-funded project
\cite{deisa}.

\subsubsection{Gluonic vertex modules}

Feynman rules $V_{G,r}$ for the gluonic vertices are implemented in \\
\verb|mod_vertex_*.F90|, where
\texttt{*} is $r$ (the number of gluons) copies of character
``\texttt{g}''. Currently vertices for $r \le 5$ have been
implemented.

The modules for $r \ge 1$ are called \verb|vertex_*| and have a
very similar structure. The top level routine is
\texttt{vert\_*(k,lorentz,colour)}, which takes as arguments the
momenta, lorentz and colour indices and returns the complete
\textit{symmetrised} Feynman rule for that vertex as a complex
number. The colour array is ignored if we are using twisted boundary
conditions. The module assumes that the monomials written by the
\hippy\ code have been partially symmetrised as described above. The
remaining symmetrisation over the distinct cosets is carried out
automatically in the top level routine. The reduced vertex is
calculated from the monomials in \verb|yvertex_*()|.

This code structure is repeated using instances of type
\texttt{taylor} instead of complex numbers, providing
analytic derivatives of the Feynman rules. The top level routine in
this case is called \texttt{taylor\_vert\_*()}. As detailed above,
coding the algorithm twice in this way is exploited in the runtime
testing of the code.

The remainder of the \texttt{mod\_vertex\_qq*.F90} files is taken up
with initialisation code (which reads in all the vertex files at
runtime) and testing routines.

Only \texttt{mod\_vertex\_gg.F90} differs from this overall structure.
This calculates the gluon propagator. In this case, the top-level
routine is \\
\texttt{gluon\_prop(k,colour)}. The propagator is
simultaneously calculated for all Lorentz index combinations, packaged
as a $D \times D$ matrix implemented as derived type \verb|mat4x4|.

The choice of gauge is controlled by variable \verb|Galpha|, specified
in runtime input file \verb|vertex_gg_parm.in|. Value \verb|1|
corresponds to Feynman gauge and \verb|0| to Landau gauge. As detailed
in Ref.~\cite{Drummond:2002yg}, the inverse propagator does not exist
in Landau gauge, so an intermediate gauge parameter \verb|Gbeta|
should be used. \verb|gauge_family = 'landau'| is the usual choice,
where the gauge correction is based on the four-vector
$\bld{k}$. Changing this to \verb|'coulomb'| uses just the spacelike
components instead, but this option is not fully tested.

As evaluation of the gluon propagator can be an expensive part of a
perturbative calculation, the propagators for various gluon actions
are hardwired in \verb|mod_vertex_gg.F90|. Their use is specified by
changing variable \verb|action| in \verb|vertex_gg_parm.in| from
\verb|'python'| to appropriate other character strings which are 
listed in subroutine \verb|vertex_gg_params()|.

A gluon mass can be added to the propagator. Its square is
\verb|gluon_mass2|, specified in \verb|vertex_gg_parm.in|.

Again, this code is repeated using automatic derivatives packaged
instead in a \verb|taylor|-valued matrix of type
\verb|tayl_mat4x4|. The remainder of the module contains runtime tests
and initialisation routines.

Associated with the gluon modules are modules calculating the Feynman
rules associated with the Fadeev-Popov ghost fields and the Haar
measure. These are implemented as hardwired formul\ae\ in
\verb|mod_vertex_hhstar.F90| (for $r \le 2$) and in
\verb|mod_vertex_meas_gg.F90| (only for $r=2$).

\subsubsection{Fermion vertex modules}

Feynman rules $V_{F,r}$ for the fermionic vertices (the
\textit{unsymmetrised} version of Eqn.~(\ref{eqn_ferm_symm})) are
calculated in the files \texttt{mod\_vertex\_qq*.F90}, where
\texttt{*} is $r$ (the number of gluons) copies of character
``\texttt{g}''. Currently vertices for $r \le 3$ have been
implemented. The modules for $r \ge 1$ are called \texttt{vertex\_qq*}
and have a very similar structure. The top level routine is
\texttt{vert\_qq*(k,lorentz,colour)}, which takes as arguments the
momenta, lorentz and colour indices and returns the complete Feynman
rule for that vertex as an instance of type \texttt{spinor}. The
convention is that \texttt{k(1:r+2)} is $(/ \bld{q}, \bld{p},
\bld{k}_1 , \dots , \bld{k}_r /)$. The colour array has the same
ordering, but is ignored if we are using twisted boundary conditions.

Internally, this function calculates the reduced vertex using
\texttt{yvertex\_qq*()} and multiplies on the 
appropriate Clebsch-Gordon factor, calculated using
\\
\texttt{mod\_colour.F90}.  

\paragraph{\texttt{yvertex\_qq*()}} 

Implements the summands and factors
decomposition described in Sec.~\ref{sec_summands_factors}. The
reduced vertices for each factor are calculated in
\\
\texttt{yvertex\_qq*\_partial()}. The sum over partitions $\sum_{P \in
P^{(r)}}$ naturally translates into a block structure for this
routine, the order of which follows that of the expressions in
Appendix~\ref{sec_concrete_factors}.

\paragraph{\texttt{yvertex\_qq*\_partial()}} 

Implements the two-level
field/algebra algorithm as per Section~\ref{sec_two_level}. The
``upper-level'', fattened, field vertices $Z_{F,r}$ are calculated in
\texttt{yvertex\_qq*\_partial\_noalg()}, combining monomials that have
been read into the module at runtime. The ``lower-level'', algebra
vertices $X_{F,r}$ are calculated in \texttt{yvertex\_qq*\_algebra()}.
Once more, the sum over partitions translates into a block structure
for the code, and the order reflects the expressions in
Appendix~\ref{sec_concrete_algebra}. As discussed earlier, for certain
Feynman diagrams we can comment out some blocks on symmetry
grounds.

\paragraph{\texttt{yvertex\_qq*\_algebra()}} 

Calculates $X_{F,r}$ in 
terms of the monomial-calculated $W_{F,r}$ from
\texttt{y\_qq*\_alg\_basic()}, as detailed in Sec.~\ref{sec_reunit}.
There is the option to apply no reunitarisation (setting $X_{F,r} =
W_{F,r}$), or the hardwired projection described in the text.
Additional projection methods, such as ``stouting''
\cite{Morningstar:2003gk},
could also be implemented here.

This code structure is repeated using instances of type
\texttt{tayl\_spinor} instead of type \texttt{spinor}, providing
analytic derivatives of the Feynman rules. The top level routine in
this case is called \texttt{taylor\_vert\_qqg*()}. As detailed above,
coding the algorithm twice in this way is exploited in the runtime
testing of the code.

The remainder of the \texttt{mod\_vertex\_qq*.F90} files is taken up
with initialisation code (which reads in all the vertex files at
runtime) and testing routines.

Only \texttt{mod\_vertex\_qq.F90} differs from this overall structure.
This calculates the fermion propagator. In this case, the top-level
routine is 
\\
\texttt{quark\_prop(k,colour)} where the momentum specified
is $\bld{q}$ (which flows in the direction of the fermion arrow).
\texttt{inv\_quark\_prop()} calculates the two-point function for the
fermions and implements the summands and factors decomposition
described in Sec.~\ref{sec_summands_factors}. The reduced vertex for
each factor is calculated in \texttt{inv\_quark\_prop\_partial()}.
This is the lowest-level routine, as there is no algebra contribution
for $r=0$. These routines are followed in the code by type
\texttt{tayl\_spinor} versions and by initialisation and tests.

\subsubsection{Runtime specification of Feynman rules}

The \hpsrc\ code allows for the use of multiple fermion types, each
with their own Feynman rules constructed in their own way. The details
of all of this are specified at runtime by the file
\texttt{vertex\_qq\_composite.in}, which is read by all the
\texttt{mod\_vertex\_qq*.F90} files.
The input takes the form of a NAMELIST. A typical example is
given in Appendix~\ref{app_composite}.

\section{Conclusions}
\label{sec_conc}

The accuracy of simulations of lattice QCD (and other field theories)
has been markedly improved by the use of Symanzik- and
radiatively-improved actions and measurement operators. In addition to
the calculation of the radiative corrections, this improvement
programme also requires a concomitant effort in calculating associated
renormalisation parameters that are vital to the continuum and chiral
extrapolations. Lattice perturbation theory is an essential tool in
these calculations. For some quantitities we can use other techniques,
in particular non-perturbative renormalisation (NPR)
\cite{Martinelli:1994ty,Sommer:2006sj},
high-$\beta$ simulations
\cite{Trottier:2001vj,Allison:2008ri}
and stochastic techniques
\cite{DiRenzo:2007qf}.
In cases where these can be used, lattice perturbation theory provides
a valuable check of their results and, in the case of high-$\beta$
simulations, important constraints on the coefficients of the low
powers of $\alpha_s$. When there is complicated mixing of operators,
which happens in a lot of physically relevant measurements,
statistical errors lead to the failure of the NPR and high-$\beta$
techniques and lattice perturbation theory is the only alternative.
Stochastic perturbation theory is also very expensive for theories
with dynamical fermions.

Lattice perturbation theory for improved actions is complicated: not
only does it include many vertices not present in the continuum, the
Feynman rules also contain a very large number of terms. These
complications vanish, of course, in the continuum limit as the lattice
spacing $a$ tends to zero. The point is, however, that lattice
perturbation theory is used to correct for the missing momentum modes
on a discrete lattice, so the presence of these complications are
central to its utility. Many of the complicating terms violate Lorentz
symmetry and are mathematically complicated. This makes analytic
evaluation of Feynman integrals impossible in almost all cases. It is
therefore crucial that we develop a robust computational method for
deriving and then using these Feynman rules in perturbation theory.
This method must be efficient as well as accurate.

In this paper we have described such a method. We have used a
two-pronged approach, with separate programs to expand the action and
derive the Feynman rules, and then to use these rules to calculate
Feynman integrals. In both cases, we have developed efficient
algorithms to minimise the computational expense. Indeed, without
these efficient algorithms we find the calculations to be impossible
for many realistic action choices. We have also presented a full,
working implementation of the algorithm in the form of the
\hippy\ and \hpsrc\ codes. The programming languages used for the
implementations are Python and Fortran95, respectively. Python
offers good list-handling and object-orientation features that are
suited to the expansion of the action to derive the Feynman
rules. Fortran95 is numerically efficient, an over-riding concern for
the evaluation (or estimation) of expensive Feynman integrals.

We stress that our method is generically applicable. In the text, we
have focused on the features of our method that make possible
calculations using realistic NRQCD and HISQ fermion actions. This bias
merely reflects the problems for which we have used the method
most so far. We have considered other actions, both by expanding them
using the \hippy\ code and by hardwiring the hand-derived Feynman
rules in the \hpsrc\ vertex modules.

Particular strengths of our approach are the ability to deal with
complicated colour and spin structures in actions and the breaking
down of actions into simpler sub-structures for faster evaluation. We
have also provided a particularly simple way of describing actions
that reduces the scope for transcription errors. The comprehensive
suite of tests in the code also give confidence in the accuracy of the
basic components of both the \hippy\ and \hpsrc\ code suites. 

As well as extending the basic functionality of the codes, e.g. to
describe actions with multiply nested levels of link fattening, the
codes are easily extended to a wide range of related problems in
lattice simulation, as has already been done for lattice chiral
perturbation theory
\cite{Borasoy:2005nz}
and Schr\"{o}dinger functional calculations
\cite{Takeda:2007dt,Takeda:2008rr}.
Another area in which the code might be used is in calculations where
the lattice gauge field is split into a fluctuating, quantum piece and
an external background field
\cite{Luscher:1995vs,Luscher:1995np}. 
The authors are happy to provide further details on the implementation
of this, as well as additional advice on the use of the \hippy\ and
\hpsrc\ codes.

\section{Acknowledgements}

A.H. thanks the U.K. Royal Society for financial support. The
University of Edinburgh is supported in part by the Scottish
Universities Physics Alliance (SUPA).  G.M.v.H. is supported by the
Deutsche Forschungsgemeinschaft in the SFB/TR~09.  This work has made
use of the resources provided by the Edinburgh Compute and Data
Facility (ECDF; \url{http://www.ecdf.ed.ac.uk}). The ECDF is partially
supported by the eDIKT initiative (\url{http://www.edikt.org.uk}). We
also acknowledge support from the DEISA Extreme Computing
Initiative (\url{http://www.deisa.eu/science/deci}).

\section{Appendices}
\appendix

\section{Explicit expressions for factors}
\label{sec_concrete_factors}

\begin{figure}
\centering
\includegraphics[height=0.9\textwidth,clip,angle=-90]{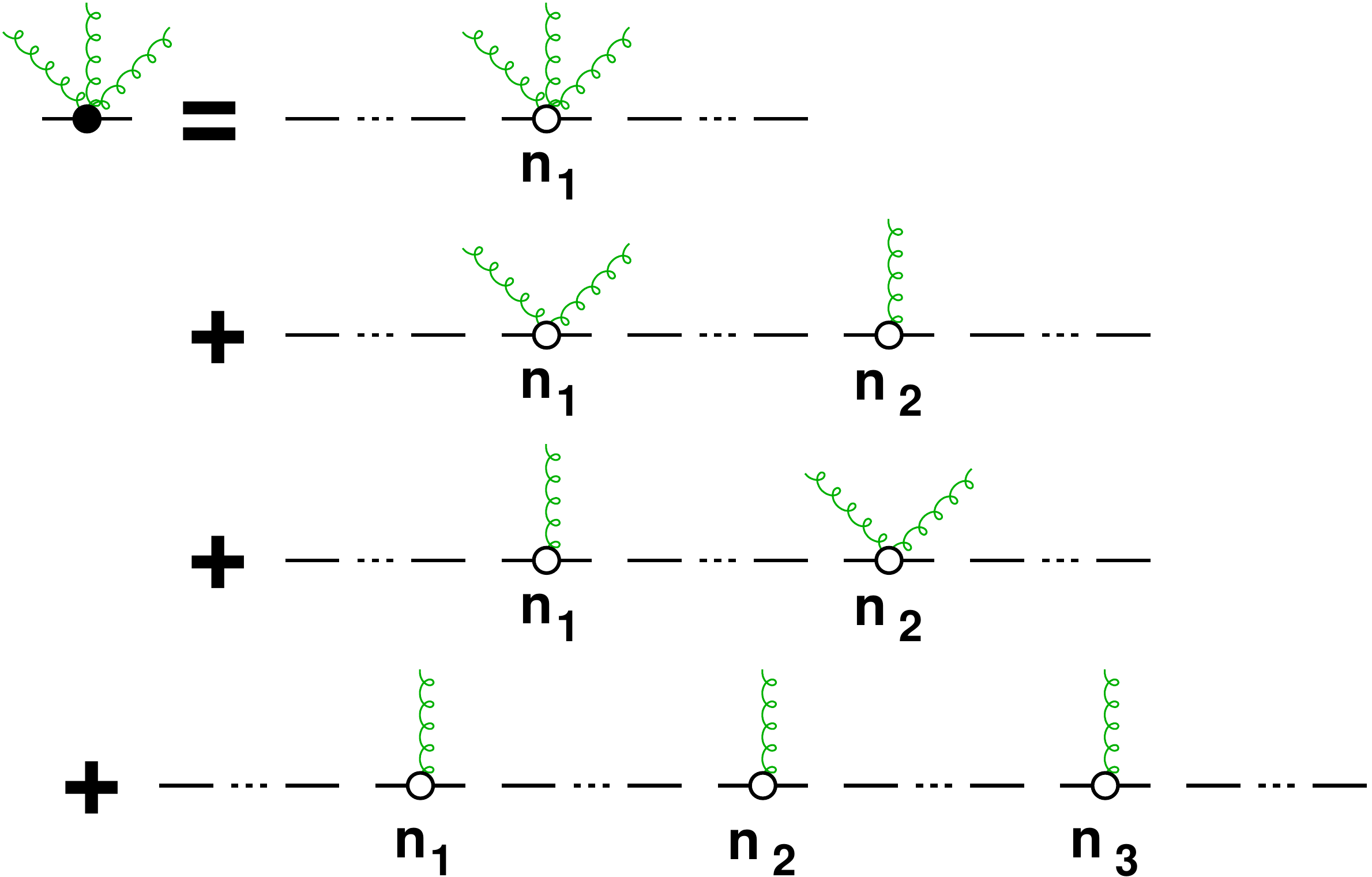}
\caption{\label{fig_summands}A graphical representation of the reduced
vertex $Y_{F,r}$ for $r=3$ for a single summand of an action that is
composed of a number of factors, as described mathematically in
Eqn.~(\ref{eqn_summands}) and implemented in the \hpsrc\ code. The
four terms in Eqn.~(\ref{eqn_summands}) are separately shown, with
$n_{1,2,3}$ showing the number of the factors from which each of the
gluons are drawn. Open circles denote vertices associated with
individual factors. Factors contributing no gluons are shown as ``$-
\cdots -$''. Momentum flow in the vertices is not shown, but can be
deduced from Eqn.~(\ref{eqn_summands}).}
\end{figure}

Using Table~\ref{tab_partitions}, we here give explicit expressions
for Eqn.~(\ref{eqn_factors}) for $Y_{F,r}$ for a summand in the action
with $N$ factors, for $r=0\dots 3$ (the range of $r$ implemented in
the \hpsrc\ code):
\begin{equation}
Y_{F,0}(\bld{p},\bld{q}) = 
\prod_{k = 1}^N
Y^{(k)}_{F,0}(\bld{p},\bld{q})
\end{equation}
\begin{multline}
Y_{F,1}(\bld{p},\bld{q};\bld{k}_1,\mu_1) =
\sum_{n_1 = 1}^N 
\left( 
\prod_{k = 1}^{n_1 - 1}
Y^{(k)}_{F,0}(\bld{p},-\bld{p})
\right)
Y^{(n_1)}_{F,1}(\bld{p},\bld{q};\bld{k}_1,\mu_1)
\\
\times
\left( 
\prod_{k = n_1 + 1}^N
Y^{(k)}_{F,0}(-\bld{q},\bld{q})
\right)
\end{multline}
\begin{multline}
Y_{F,2}(\bld{p},\bld{q};\bld{k}_1,\mu_1;\bld{k}_2,\mu_2) =
\sum_{1 \le n_1 \le N} 
\left( 
\prod_{k = 1}^{n_1 - 1}
Y^{(k)}_{F,0}(\bld{p},-\bld{p})
\right)
\\
\times
Y^{(n_1)}_{F,2}(\bld{p},\bld{q};\bld{k}_1,\mu_1;\bld{k}_2,\mu_2)
\left( 
\prod_{k = n_1 + 1}^N
Y^{(k)}_{F,0}(-\bld{q},\bld{q})
\right)
\\
+
\sum_{1 \le n_1 < n_2 \le N} 
\left( 
\prod_{k = 1}^{n_1 - 1}
Y^{(k)}_{F,0}(\bld{p},-\bld{p})
\right)
Y^{(n_1)}_{F,1}(\bld{p},-\bld{p}-\bld{k}_1;\bld{k}_1,\mu_1)
\\
\times
\left( 
\prod_{k = n_1+1}^{n_2-1}
Y^{(k)}_{F,0}(\bld{p}+\bld{k}_1,-\bld{p}-\bld{k}_1)
\right)
\\
\times
Y^{(n_2)}_{F,1}(\bld{p}+\bld{k}_1,\bld{q};\bld{k}_2,\mu_2)
\left( 
\prod_{k = n_2 + 1}^N
Y^{(k)}_{F,0}(-\bld{q},\bld{q})
\right)
\end{multline}
\begin{multline}
Y_{F,3}(\bld{p},\bld{q};\bld{k}_1,\mu_1;\bld{k}_2,\mu_2;
\bld{k}_3,\mu_3) =
\\
\sum_{1 \le n_1 \le N} 
\left( 
\prod_{k = 1}^{n_1 - 1}
Y^{(k)}_{F,0}(\bld{p},-\bld{p})
\right)
Y^{(n_1)}_{F,3}(\bld{p},\bld{q};\bld{k}_1,\mu_1;\bld{k}_2,\mu_2;
\bld{k}_3,\mu_3)
\\
\times
\left( 
\prod_{k = n_1 + 1}^N
Y^{(k)}_{F,0}(-\bld{q},\bld{q})
\right)
\\
+
\sum_{1 \le n_1 < n_2 \le N} 
\left( 
\prod_{k = 1}^{n_1 - 1}
Y^{(k)}_{F,0}(\bld{p},-\bld{p})
\right)
Y^{(n_1)}_{F,2}(\bld{p},\bld{q}+\bld{k}_3;\bld{k}_1,\mu_1;
\bld{k}_2,\mu_2)
\\
\times
\left( 
\prod_{k = n_1+1}^{n_2-1}
Y^{(k)}_{F,0}(-\bld{q}-\bld{k}_3,\bld{q}+\bld{k}_3)
\right)
\\
\times
Y^{(n_2)}_{F,1}(-\bld{q}-\bld{k}_3,\bld{q};\bld{k}_3,\mu_3)
\left( 
\prod_{k = n_2 + 1}^N
Y^{(k)}_{F,0}(-\bld{q},\bld{q})
\right)
\\
+
\sum_{1 \le n_1 < n_2 \le N} 
\left( 
\prod_{k = 1}^{n_1 - 1}
Y^{(k)}_{F,0}(\bld{p},-\bld{p})
\right)
Y^{(n_1)}_{F,1}(\bld{p},-\bld{p}-\bld{k}_1;\bld{k}_1,\mu_1)
\\
\times
\left( 
\prod_{k = n_1+1}^{n_2-1}
Y^{(k)}_{F,0}(\bld{p}+\bld{k}_1,-\bld{p}-\bld{k}_1)
\right)
\\
\times
Y^{(n_2)}_{F,2}(\bld{p}+\bld{k}_1,\bld{q};\bld{k}_2,\mu_2;\bld{k}_3,\mu_3)
\left( 
\prod_{k = n_2 + 1}^N
Y^{(k)}_{F,0}(-\bld{q},\bld{q})
\right)
\\
+
\sum_{1 \le n_1 < n_2 < n_3 \le N} 
\left( 
\prod_{k = 1}^{n_1 - 1}
Y^{(k)}_{F,0}(\bld{p},-\bld{p})
\right)
Y^{(n_1)}_{F,1}(\bld{p},-\bld{p}-\bld{k}_1;\bld{k}_1,\mu_1)
\\
\times
\left( 
\prod_{k = n_1+1}^{n_2-1}
Y^{(k)}_{F,0}(\bld{p}+\bld{k}_1,-\bld{p}-\bld{k}_1)
\right)
\\
\times
Y^{(n_2)}_{F,1}(\bld{p}+\bld{k}_1,\bld{q}+\bld{k}_3;
\bld{k}_2,\mu_2)
\left( 
\prod_{k = n_2 + 1}^{n_3-1}
Y^{(k)}_{F,0}(-\bld{q}-\bld{k}_3,\bld{q}+\bld{k}_3)
\right)
\\
\times
Y^{(n_3)}_{F,1}(-\bld{q}-\bld{k}_3,\bld{q};\bld{k}_3,\mu_3)
\left( 
\prod_{k = n_3 + 1}^N
Y^{(k)}_{F,0}(-\bld{q},\bld{q})
\right)
\label{eqn_summands}
\end{multline}
The reduced vertex from the $n^\text{th}$ is denoted $Y^{(n)}_{F,r}$.
The formula for $r=3$ is illustrated graphically in
Fig.~\ref{fig_summands}.

\section{Explicit expressions for two-level actions}
\label{sec_concrete_algebra}

To make things more concrete, we give explicit formul\ae\ for the
$r=1$,~$r=2$ and~$r=3$ reduced vertices for a two-level action as
defined in Eqn.~(\ref{eqn_algebra}). This is the range of $r$
implemented in the \hpsrc\ code. Using the partitions in
Table~\ref{tab_partitions} we obtain:
\begin{equation}
Y_{F,1}(\bld{p},\bld{q}; \bld{k}_1,\mu_1) = 
\sum_{\nu_1} Z_{F,1}(\bld{p},\bld{q}; \bld{k}_1,\nu_1)
X^{\nu_1}_{F,1}(\bld{k}_1,\mu_1)
\label{eqn_Y_F1}
\end{equation}
\begin{multline}
Y_{F,2}(\bld{p},\bld{q}; \bld{k}_1,\mu_1; \bld{k}_2,\mu_2) = 
\sum_{\nu_1} Z_{F,1}(\bld{p},\bld{q}; \bld{k}_1 + \bld{k}_2,\nu_1)
X^{\nu_1}_{F,2}(\bld{k}_1,\mu_1; \bld{k}_2,\mu_2)
\\
+ \sum_{\nu_1,\nu_2} Z_{F,2}(\bld{p},\bld{q};
\bld{k}_1,\nu_1; 
\bld{k}_2,\nu_2)
X^{\nu_1}_{F,1}(\bld{k}_1,\mu_1)
X^{\nu_2}_{F,1}(\bld{k}_2,\mu_2)
\label{eqn_Y_F2}
\end{multline}
\begin{multline}
Y_{F,3}(\bld{p},\bld{q}; \bld{k}_1,\mu_1; \bld{k}_2,\mu_2; 
\bld{k}_3,\mu_3) = 
\\
\sum_{\nu_1} Z_{F,1}(\bld{p},\bld{q}; 
\bld{k}_1 + \bld{k}_2 + \bld{k}_3,\nu_1)
X^{\nu_1}_{F,3}(\bld{k}_1,\mu_1; \bld{k}_2,\mu_2; 
\bld{k}_3,\mu_3)
\\
+ \sum_{\nu_1,\nu_2} Z_{F,2}(\bld{p},\bld{q};
\bld{k}_1 + \bld{k}_2,\nu_1; 
\bld{k}_3,\nu_2)
X^{\nu_1}_{F,2}(\bld{k}_1,\mu_1; \bld{k}_2,\mu_2)
X^{\nu_2}_{F,1}(\bld{k}_3,\mu_3)
\\
+ \sum_{\nu_1,\nu_2} Z_{F,2}(\bld{p},\bld{q};
\bld{k}_1,\nu_1; 
\bld{k}_2 + \bld{k}_3,\nu_2)
X^{\nu_1}_{F,1}(\bld{k}_1,\mu_1)
X^{\nu_2}_{F,2}(\bld{k}_2,\mu_2; \bld{k}_3,\mu_3)
\\
+ \sum_{\nu_1,\nu_2,\nu_3} Z_{F,3}(\bld{p},\bld{q};
\bld{k}_1,\nu_1; 
\bld{k}_2,\nu_2;\bld{k}_3,\nu_3)
X^{\nu_1}_{F,1}(\bld{k}_1,\mu_1)
X^{\nu_2}_{F,1}(\bld{k}_2,\mu_2)
X^{\nu_3}_{F,1}(\bld{k}_3,\mu_3)
\label{eqn_Y_F3}
\end{multline}

\section{Example of \texttt{vertex\_qq\_composite.in}}
\label{app_composite}

Here we provide an explicit example of the runtime file used to
specify the Feynman rules used by the \hpsrc\ code.

The calculation will use two types of quarks: relativistic HISQ and
heavy NRQCD. It is assumed that the appropriate vertex and, where
relevant, algebra files have been precomputed using the \hippy\ code
and stored in the locations specified in the \hpsrc\ file
\verb|paths.in|.

\begin{verbatim}
&vertex_qq_composite
        no_quark_types = 2
! First quark type: HISQ
!   All RH indices set to 1 to denote first quark type
        summands(1) = 1
        factors(1:1,1) = 1
        summand_amps(1:1,1) = 1.d0
        opname(1:1,1,1) = "asq_for_hisq_"
        algname(1) = "Ffat7"
	reunit_to_apply_to_alg(1) = "project"
! Second quark type: NRQCD
!   All RH indices set to 2 to denote second quark type
        summands(2) = 2
        factors(1:2,2) = 0,4
        summand_amps(1:2,2) = 1.d0,-1.d0
        opname(1:4,2,2) = "nrqcd_A","nrqcd_B","nrqcd_C","nrqcd_A"
        algname(2) = "simple"
	reunit_to_apply_to_alg(2) = "none"
! N.B. no opname(:,1,2) definition because first summand has no
!   HiPPy inputs because it is a constant
/
\end{verbatim}

That there are two quark types is specified in the NAMELIST using
\\
\verb|no_quark_types|. Within the \hpsrc\ code we control which set of
Feynman rules we use by setting variable \verb|quark_type| to~1 or~2 
prior to calling 
\\
\verb|quark_prop()| or \verb|vert_qqg*()|.

The first fermions are relativistic HISQ quarks. There is no splitting
of the action into summands and factors, so \texttt{summands(qt)=1}
and 
\\
\texttt{factors(1:summands(qt),qt)=1}. Similarly, the only
summand has amplitude \texttt{summand\_amps(1:summands(qt),qt)=1.0d0}.
The files used to construct $Z_{F,r}$ for each factor for each summand
are assumed to have been generated by the \hippy\ code and stored in
files named \texttt{vertex\_\%sqq*.in} where \texttt{\%s} is replaced
by appropriate text for each factor for each summand as specified in
\texttt{opname(ft,sm,qt)}, where \texttt{ft} runs from 1 to
\texttt{factors(sm,qt)} and \texttt{sm} runs from 1 to
\texttt{summands(qt)}. In this case, the only vertex files are those
with \texttt{\%s} replaced by \verb|asq_for hisq_|. 

The name of the algebra file is given by \verb|algname(qt)| (note
there is no trailing underscore in this case). We specify that
hardwired projection is required to relate $W_{F,r}$ in
Eqn.~(\ref{eqn_wy}) to $X_{F,r}$ in Eqn.~(\ref{eqn_xy}) using
\verb|reunit_to_apply_to_alg(qt)|. The \hpsrc\ code currently assumes
that the same algebra files are used for all factors and summands of a
particular quark type. Again, it is assumed that the files defining
these have been pre-produced using the \hippy\ code and stored in
files named \texttt{algebra\_\%s\%i\_qq*.in}. \texttt{\%s} is replaced
by \texttt{algname(qt)} and \texttt{\%i} runs from 0 to $D-1$, where
$D$ is the number of dimensions.

In this explanation, we have used \texttt{qt} to denote the quark
type, and other intuitive labels to specify array sizes.
Unfortunately, NAMELISTs only support numbered array entries.

The second block of the NAMELIST defines an NRQCD action of the
heuristic form $\bar{\psi} (1-ABCA) \psi$. There are two summands, the
first with 0 factors (which gives default answer 1) and the second
with 4. The minus sign in front of the second summand is specified in
\texttt{summand\_amps}. The first summand has no factors, so
\texttt{opname} is not specified. The second summand has 4 factors,
and the filenames are specified. Note that the first and last
filenames are the same. The algebra is specified as before, and no
hardwired reunitarisation is required.

In both cases, the correct spin algebra is specified in the headers of
the input files written by the \hippy\ code.

%
%
%
%
%


\end{document}